\documentclass[twocolumn,prd,showpacs]{revtex4}
\usepackage{graphicx}
\usepackage{textcomp,amsmath,amssymb}

\begin{document}

\title{Modifications for numerical stability of black hole evolution}

\author{Hwei-Jang Yo${}^{1}$}\email{hjyo@phys.ncku.edu.tw}
\author{Chun-Yu Lin${}^{2}$}
\author{Zhoujian Cao${}^{3}$} \email{zjcao@amt.ac.cn}

\affiliation{${}^{1}$Department of Physics, National Cheng-Kung University, Tainan 701, Taiwan, \\
${}^{2}$National Center for High-Performance Computing, Hsinchu 300, Taiwan\\
${}^3$Institute of Applied Mathematics, Academy of Mathematics and Systems Science, 
Chinese Academy of Sciences, Beijing 100190, China}

\date{\today}

\begin{abstract}
We experiment with several new modifications to the
Baumgarte-Shapiro-Shibata-Nakamura (BSSN) formulation of Einstein's field
equation, and demonstrate how these modifications affect the stability of
numerical black hole evolution.
With these modifications, we obtain accurate and stable simulations of 
both single excised Kerr-Schild black holes and punctured binary black holes.
\end{abstract}
\pacs{04.25.Dm, 04.30.Db, 95.30.Sf, 97.60.Lf}

\maketitle

\section{Introduction}

Numerical relativity is aimed at solving Einstein's equations with
the aid of computers.
It took decades to reformulate Einstein's equations to
the required stability and accuracy in simulations.
Breakthroughs in 2005 and 2006 \cite{pref05,NR06} brought this
development to a more mature status,
gave confidence to the community in modeling gravitational wave
sources and extracting information of gravitation radiation from the
simulations of binary compact objects.
Since then, observations in simulations which have been refined and extensively
studied over the past few years include long term gravitational waves and
the final state of the binary compact objects merger (see review articles
\cite{Lehner_reviewNR01,reviewBBH,reviewBHNS} and reference therein),
gravitational recoil of a black hole \cite{BHkicks}, relativistic
jet formation from mergers of black holes \cite{JetBH}, and neutron stars
\cite{JetNS}.
These also provide new insights into mathematical general relativity.
Numerical relativity has now become an indispensable and efficient tool in the
research of general relativity and astrophysics.

Among the numerous reformulations of Einstein's equations,
two particular formulations are most frequently adopted for the simulations of black holes. One
of them is the generalized harmonic formulation, in either second-order
formulation \cite{pref05}, or fully first-order formulation \cite{lial06}
(In either case the key ingredient for stability of this formulation
is a constraint damping mechanism \cite{bral99}). The other is
the Baumgarte--Shapiro--Shibata--Nakamura (BSSN) system
\cite{BSSN95_99}, which has been implemented by many groups using
finite difference codes, in first-order-in-time and
second-order-in-space form.

There have been many studies in modifying the BSSN formulation to increase its
numerical stability.
For example, Alcubierre and Br\"ugmann \cite{almb01} combined the BSSN
formulation with constraints enforcing the traceless
condition of the conformal extrinsic curvature in every time step,
replacing the conformal connection function ${\tilde\Gamma}^i$ with that
calculated from the Christoffel symbols whenever ${\tilde\Gamma}^i$ is
undifferentiated.
With these modifications, they evolved single black holes in Kerr-Schild
coordinates stably with octant grid symmetry, and all fields settled down to
equilibrium without encountering any instability.
The confinement to grid symmetry was relaxed later in \cite{yhbs02}, by adding
the $\Gamma$-constraint to the field equation of ${\tilde\Gamma}^i$ in order
to suppress instability, as well as by employing alternative techniques
in the enforcement of other algebraic constraints.
Yoneda and Shinkai analyzed analytically the constraint propagation of BSSN
formulation in \cite{ygsh02}, and proposed adjustments to obtain better
stability by adding constraint terms.
The follow-up work in \cite{kksh08} numerically verified the advantage of
their adjustments in stability over the original BSSN formulation.
Higher-order derivatives of constraint was also added to the field
equation in \cite{ttys12} to enhance stability.
A first-order BSSN formulation has been developed in \cite{bjet12} to seek more
stable performance in simulations.

In this work, we report numerical tests of new modifications
to the BSSN system. The work can be considered as an extended study
based on an earlier article \cite{yhbs02}. The modifications include
(1) adding the $\Gamma$-constraint to the field equation of the
conformal three-metric; (2) replacing the conformal connection
function calculated from the conformal metric, i.e.,
${\tilde\Gamma}^i_{\bf g}$, with the independent
${\tilde\Gamma}^i$, and enhancing the derivative of the unimodular
determinant constraint in the connection with an irreducible
decomposition; (3) applying, with some deformation, the adjustment of the field equation of
the conformal extrinsic curvature with the momentum constraint
proposed in \cite{ygsh02,kksh08}. We also
emphasize the alternative method in \cite{yhbs02}
in the enforcement of the unimodular constraint and the traceless
conformal extrinsic curvature constraint in every timestep. We
experiment with these modifications on simulations of single
Kerr-Schild black hole, and obtain evolutions with
long-term stability. These modifications, also applied to the
evolutions of binary punctured black hole, demonstrate better stability and
accuracy than the original BSSN formulation against numerical errors,
either from finite-differencing or from mesh refinement.

The paper is organized as follows: We summarize the BSSN formulation in
Sec.~II. Our modifications of the BSSN scheme are described in Sec.~III.
Single black hole spacetimes in Kerr-Schild coordinates and binary black hole
with punctures are described in Sec.~IV.
We discuss numerical implementations and the gauge conditions in Sec.~V.
In Sec.~VI we present results of our simulations for both single black hole
and binary black holes.
We summarize and discuss the implications of our findings in Sec.~VII.
Throughout this paper we adopt geometric units, $G=c=1$.
\section{The BSSN formulation}
\label{Bf}
The metric in the ADM form is
\begin{equation} \label{adm_metric}
{\rm d}s^2=-\alpha^2{\rm d}t^2+\gamma_{ij}({\rm d}x^i+\beta^i{\rm d}t)
({\rm d}x^j+\beta^j{\rm d}t),
\end{equation}
wherein $\alpha$ is the lapse function, $\beta^i$ is the shift vector,
and $\gamma_{ij}$ is the spatial three-metric.  Throughout this paper, Latin
indices are spatial indices and run from 1 to 3, whereas Greek indices
are space-time indices and run from 0 to 3.

Einstein's equations can then be decomposed into the Hamiltonian
constraint ${\mathcal H}$ and the momentum constraints ${\mathcal
M}_i$
\begin{align}
   {\mathcal H} &\equiv R - K_{ij}K^{ij} + K^2 = 0, \label{ham1} \\
   {\mathcal M}_i&\equiv \nabla_j K^{j}_{~i} - \nabla_i K = 0, \label{mom1}
\end{align}
and the evolution equations
\begin{align}
\frac{\rm d}{{\rm d}t} \gamma_{ij} & =- 2 \alpha K_{ij},\label{gdot1} \\
\frac{\rm d}{{\rm d}t}K_{ij}&=-\nabla_i \nabla_j\alpha+\alpha( R_{ij}
        - 2 K_{i\ell} K^\ell_{~j} + K K_{ij}).       \label{Kdot1}
\end{align}
Here we have assumed vacuum $T_{\alpha\beta} = 0$ and have used
\begin{equation}
\frac{\rm d}{{\rm d}t} = \frac{\partial}{\partial t} - \pounds_{\vec\beta},
\end{equation}
where $\pounds_{\vec\beta}$ is the Lie derivative with respect to
$\beta^i$.   $\nabla_i$ is the covariant derivative associated with
$\gamma_{ij}$, $R_{ij}$ is the three-dimensional Ricci tensor
\begin{align} \label{ricci}
   R_{ij}=&\frac{1}{2} \gamma^{k\ell} \left( \gamma_{kj,i\ell}
    +\gamma_{i\ell,kj}-\gamma_{k\ell,ij}-\gamma_{ij,k\ell}\right)\nonumber\\
   &+ \gamma^{k\ell} \left( \Gamma^m{}_{i\ell}
    \Gamma_{mkj} - \Gamma^m{}_{ij} \Gamma_{mk\ell} \right),
\end{align}
where
\begin{equation}
\Gamma^{i}{}_{jk}\equiv\frac{1}{2}\gamma^{i\ell}(\gamma_{\ell j,k}+
\gamma_{\ell k,j}-\gamma_{jk,\ell}).
\end{equation}
And $R$ is its trace $R = \gamma^{ij} R_{ij}$.

In the BSSN formalism \cite{BSSN95_99}, the above ADM equations are
rewritten by introducing the conformally related metric $\tilde
\gamma_{ij}$
\begin{equation}\label{confgij}
\tilde \gamma_{ij} = e^{- 4 \phi} \gamma_{ij},
\end{equation}
with the conformal exponent $\phi$ chosen so that the determinant
$\tilde\gamma$ of $\tilde \gamma_{ij}$ is unity
\begin{equation}
e^{4 \phi} = \gamma^{1/3},
\end{equation}
where $\gamma$ is the determinant of $\gamma_{ij}$.
The traceless part of the extrinsic curvature $K_{ij}$, defined by
\begin{equation}\label{tracelessK}
A_{ij} = K_{\langle ij\rangle}\equiv K_{ij} - \frac{1}{3} \gamma_{ij} K,
\end{equation}
where $K_{ij}$ with two indices between $\langle\rangle$ is to take the
symmetric and traceless part of $K_{ij}$, and
$K = \gamma^{ij} K_{ij}$ is the trace of the extrinsic curvature,
is conformally decomposed according to
\begin{equation}
\tilde A_{ij} = e^{- 4 \phi} A_{ij}.
\end{equation}
The conformal connection functions $\tilde{\Gamma}^i$, initially defined as
\begin{equation} \label{Gamma}
\tilde{\Gamma}^i \equiv \tilde{\gamma}^{jk} \tilde{\Gamma}^i_{jk}
        = - \tilde{\gamma}^{ij}_{~~,j},
\end{equation}
are regarded as independent variables in this formulation.

The evolution equations of BSSN formulation can be written as
\begin{align}
\frac{\rm d}{{\rm d}t}\phi&=-\frac{1}{6}\alpha K,\label{eq:evolphi}\\
\frac{\rm d}{{\rm d}t}\tilde{\gamma}_{ij}&=-2\alpha\tilde{A}_{ij},\label{dtg}\\
\frac{\rm d}{{\rm d}t}K&=\alpha\left(\tilde{A}_{ij}\tilde{A}^{ij}
+\frac{1}{3}K^2\right)-\nabla^2\alpha, \label{dtK}\\
\frac{\rm d}{{\rm d}t}\tilde{A}_{ij}&=
        \alpha(K\tilde{A}_{ij}-2\tilde{A}_{ik}\tilde{A}^k{}_j)+
e^{-4\phi}(\alpha R_{\langle ij\rangle}-\nabla_{\langle i}\nabla_{j\rangle}
\alpha),\label{dtA}\\
\partial_t\tilde{\Gamma}^i&=2\alpha\left(\tilde{\Gamma}^i_{jk}
    \tilde A^{jk}-\frac{2}{3}\tilde{\gamma}^{ij}K_{,j}
    +6\tilde{A}^{ij}\phi_{,j}\right)-2\tilde{A}^{ij}\alpha_{,j}\nonumber\\
   +&\beta^j\tilde{\Gamma}^i{}_{,j}-\tilde{\Gamma}^j\beta^i{}_{,j}
+\frac{2}{3}\tilde{\Gamma}^i\beta^j{}_{,j}
+\tilde{\gamma}^{jk}\beta^i{}_{,jk}+
  \frac{1}{3}\tilde{\gamma}^{ij}\beta^k{}_{,jk}.\label{dtGamma}
\end{align}
The Ricci tensor $R_{ij}$ can be written as a sum of two pieces
\begin{equation}
   R_{ij} = \tilde{R}_{ij} + R^{\phi}_{ij},
\end{equation}
where $R^{\phi}_{ij}$ is given by
\begin{equation}
   R^{\phi}_{ij}=-2\tilde{\nabla}_i\tilde{\nabla}_j\phi-2\tilde{\gamma}_{ij}
    \tilde{\nabla}^2\phi+4\tilde{\nabla}_i\phi\tilde{\nabla}_j\phi
   -4\tilde{\gamma}_{ij}\tilde{\nabla}^k\phi\tilde{\nabla}_k\phi,
\end{equation}
where ${\tilde\nabla}_i$ is the covariant derivative with respect to 
${\tilde\gamma}_{ij}$,
while, with the help of the $\tilde \Gamma^i$, $\tilde{R}_{ij}$
can be expressed as
\begin{align}
   \tilde{R}_{ij}=&-\frac{1}{2}\tilde{\gamma}^{mn}\tilde{\gamma}_{ij,mn}
     +\tilde{\gamma}_{k(i}\tilde{\Gamma}^k{}_{,j)}
     +\tilde{\Gamma}^k\tilde{\Gamma}_{(ij)k} \nonumber \\
   &+\tilde{\gamma}^{mn}[2\tilde{\Gamma}^k{}_{m(i}\tilde{\Gamma}_{j)kn}
     +\tilde{\Gamma}^k{}_{in}\tilde{\Gamma}_{kmj}].\label{confricci}
\end{align}
The new variables are tensor densities, so that their Lie derivatives are
\begin{align}
   \pounds_{\vec\beta}K&=\beta^kK_{,k},\\
   \pounds_{\vec\beta}\phi&=\beta^k\phi_{,k} + \frac{1}{6}\beta^k{}_{,k},\\
   \pounds_{\vec\beta}\tilde{\gamma}_{ij}&=\beta^k\tilde{\gamma}_{ij,k}
 +2\tilde{\gamma}_{k(i}\beta^k{}_{,j)}-\frac{2}{3}\tilde{\gamma}_{ij}
  \beta^k{}_{,k},\label{lieg}\\
   \pounds_{\vec\beta}\tilde{A}_{ij}&=\beta^k\tilde{A}_{ij,k}
    +2\tilde{A}_{k(i}\beta^k{}_{,j)}-\frac{2}{3}\tilde{A}_{ij}\beta^k{}_{,k}.
\end{align}
The Hamiltonian and momentum constraints (\ref{ham1}) and (\ref{mom1})
can be rewritten as
\begin{align}
&{\mathcal H}=e^{-4\phi}(\tilde{R}-8\tilde{\nabla}^2\phi
  -8\tilde{\nabla}^i\phi\tilde{\nabla}_i\phi)
   +\frac{2}{3}K^2-\tilde{A}_{ij}\tilde{A}^{ij}=0,\label{ham2}\\
&{\mathcal M}_i=\tilde{\nabla}_j\tilde{A}_i{}^j+6\phi_{,j}\tilde{A}_i{}^j-
        \frac{2}{3}K_{,i}=0,\label{mom2}
\end{align}
where $\tilde{R}=\tilde{\gamma}^{ij}\tilde{R}_{ij}$.
Besides being used to obtain the evolution
equations (\ref{dtK}) and (\ref{dtGamma}) in the BSSN formulation,
the Hamiltonian and the momentum constraints are also applied to
the volume integrals of the ADM mass and the angular momentum (in vacuum),
respectively \cite{czyy08}:
\begin{align}
M=&\frac{1}{16\pi}\oint_{\partial\Omega}({\tilde\Gamma}^i-8{\tilde\gamma}^{ij}
\partial_je^\phi){\rm d}{\tilde\Sigma}_i\label{surfmass}\\
=&\frac{1}{16\pi}\int_\Omega[e^{5\phi}(\tilde{A}_{ij}\tilde{A}^{ij}-
\frac{2}{3}K^2)+\tilde{\Gamma}^j{}_{,j}-e^\phi\tilde{R}]{\rm d}^3x,
\label{volmass}\\
   J_i=&\frac{1}{8\pi} \epsilon_{ij}{}^k \oint_{\partial\Omega}
     e^{6\phi}x^j \tilde{A}^\ell{}_k{\rm d}\tilde{\Sigma}_\ell\label{surfang}\\
    =&\frac{1}{8\pi}\epsilon_{ij}{}^k\int_\Omega
            e^{6\phi}(\tilde{A}^j{}_k+\frac{2}{3}x^jK_{,k}-\frac{1}{2} x^j
     \tilde{A}_{\ell m}\tilde{\gamma}^{\ell m}{}_{,k}){\rm d}^3x,\label{volang}
\end{align}
where ${\rm d}\tilde{\Sigma}_i=(1/2)\epsilon_{ijk}{\rm d}x^j{\rm
d}x^k$. These two global quantities are useful tools for the system
diagnostics to validate the calculations.
\section{Adjusting the BSSN equations}\label{mar}
The BSSN formulation has been described in detail in previous papers
\cite{yhbs02}.
We will discuss here only the new improvements.
For a solution of the BSSN equations to be equivalent with a solution
of the ADM equations, the new auxiliary variables have to satisfy new
constraint equations.
In particular, the determinant of the conformally related metric
$\tilde\gamma_{ij}$ has to be unity,
\begin{equation} \label{detg1}
{\mathcal D}\equiv\tilde\gamma-1=0,
\end{equation}
and $\tilde A_{ij}$ has to be traceless
\begin{equation} \label{trA0}
{\mathcal T} \equiv \tilde{\gamma}^{ij}\tilde{A}_{ij} = 0,
\end{equation}
and the conformal connection functions $\tilde \Gamma^i$ have to
satisfy the identity
\begin{equation}
{\mathcal G}^i \equiv \tilde{\Gamma}^i-\tilde{\Gamma}^i_{\bf g}=
        \tilde{\Gamma}^i+\tilde\gamma^{ij}{}_{,j} = 0,\label{Gofg}
\end{equation}
where $\tilde{\Gamma}^i_{\bf g}\equiv\tilde{\gamma}^{jk}
\tilde{\Gamma}^i{}_{jk}$.
These conditions (\ref{detg1})-(\ref{Gofg}) are also viewed as constraints in
the BSSN formulation, in addition to the Hamiltonian and momentum constraints.
It is worth mentioning that in the recent efforts on the formalism extending
the solution space of Einstein's equation \cite{Z3Z4}, 
the ${\mathcal G}^i$'s are related to new dynamical variables whose evolution
is mainly driven by momentum constraint, and therefore can be regarded as the
cumulated effect of momentum constraint violations.

In an unconstrained evolution calculation, the constraints are
monitored only as a code check. It has been proven to be advantageous,
however, either to
enforce at least some of the constraints during the evolution,
or to add evolution constraint equations to the evolution equations.
\subsection{Enforcement of the constraints $\mathcal D$ and $\mathcal T$}
In the conventional adjustments \cite{almb01,LaPS02}, the algebraic
constraints (\ref{detg1}) and (\ref{trA0}) are enforced actively by replacing
$\tilde\gamma_{ij}$ and $\tilde A_{ij}$ with the following:
\begin{equation}
\tilde\gamma_{ij}\rightarrow\tilde\gamma^{-1/3}\tilde\gamma_{ij},\quad
\tilde A_{ij}\rightarrow\tilde A_{\langle ij\rangle},\label{Agave}
\end{equation}
after every time step.
With Eq.~(\ref{Agave}), the noise from the violation of
constraints (\ref{detg1}) and (\ref{trA0}) is scaled/subtracted ``evenly''
from each conformal metric/extrinsic curvature component.
These two adjustments are widely used in most of the numerical relativity
groups.

The alternative adjustments for constraints (\ref{detg1}) and (\ref{trA0}) in
\cite{yhbs02} are as follows:
Instead of treating all components of ${\tilde\gamma}_{ij}$ equally, only five
of the six components of ${\tilde\gamma}_{ij}$ need to be evolved dynamically,
and the remaining one can simply be computed using Eq.~(\ref{detg1}).
For example, let $\tilde\gamma_{zz}$ be the chosen component. Then
\begin{equation}
{\tilde\gamma}_{zz}=\frac{1+{\tilde\gamma}_{yy}{\tilde\gamma}_{xz}^2
-2{\tilde\gamma}_{xy}{\tilde\gamma}_{yz}{\tilde\gamma}_{xz}
+{\tilde\gamma}_{xx}{\tilde\gamma}_{yz}^2}{{\tilde\gamma}_{xx}
{\tilde\gamma}_{yy}-{\tilde\gamma}_{xy}^2},\label{gzz}
\end{equation}
where $\tilde\gamma_{xx}$, $\tilde\gamma_{yy}$, $\tilde\gamma_{xy}$,
$\tilde\gamma_{yz}$, $\tilde\gamma_{xz}$ are evolved with the field equation
(\ref{dtg}).
In principle, any one of these six components of $\tilde\gamma_{ij}$ can be
chosen to be computed using
Eq.~(\ref{detg1}), leaving the other five to be evolved with Eq.~(\ref{dtg}).
However, there will be extra difficulty if any of the three off-diagonal
variables is chosen since Eq.~(\ref{detg1}) gives a quadratic equation for
an off-diagonal component instead of a linear equation for a diagonal component.
Similarly, only five of the six components of ${\tilde A}_{ij}$ need to be
evolved dynamically, and the remaining one can simply be computed using
Eq.~(\ref{trA0}).
For example, let $\tilde A_{yy}$ be the chosen component. Then
\begin{equation}
{\tilde A}_{yy}=-\frac{{\tilde A}_x{}^x+{\tilde A}_z{}^z
+{\tilde A}_{xy}{\tilde\gamma}^{xy}+{\tilde A}_{yz}{\tilde\gamma}^{yz}}
{{\tilde\gamma}^{yy}}.\label{Ayy}
\end{equation}
Although any one of these six components can be chosen to be computed using
Eq.~(\ref{trA0}), leaving the other five to be evolved with Eq.~(\ref{dtA}).
However, it is not recommended to choose any of the three off-diagonal
components since the corresponding denominators for the off-diagonal
components, i.e., $\tilde\gamma^{xy}$, $\tilde\gamma^{yz}$, and
$\tilde\gamma^{xz}$, could vanish anywhere.

One of the features of the alternative adjustments on $\tilde\gamma_{ij}$ and
$\tilde A_{ij}$ is the economy compared with conventional methods.
Only five, instead of six, components for both $\tilde\gamma_{ij}$ and
$\tilde A_{ij}$ need to be evolved dynamically,
and the remaining one is determined by the algebraic constraint.
Meanwhile, the alternative adjustments ``correct'' only one component instead
of all the components.
It can be expected that results with the alternative adjustments will be more
convergent than those with conventional adjustments.
On the other hand, the obvious shortcoming for the alternative adjustments is
the asymmetric treatment of the six components.
However, as far as the cases we have ever checked, the effect of the asymmetry
is negligible.
It is helpful numerically to choose a different diagonal pair for
$\tilde\gamma_{ij}$
and $\tilde A_{ij}$, instead of the same diagonal pair, in eqns~(\ref{gzz})
and (\ref{Ayy}) to increase the
asymmetry in the evolution equations and to suppress the possible
growth of the unstable modes ignited by numerical error.
In the work we choose the pair ($\tilde\gamma_{zz},\tilde A_{yy}$),
instead of ($\tilde\gamma_{zz},\tilde A_{zz}$) in \cite{yhbs02}, to
fulfill this requirement.
\subsection{Decomposition}\label{dp}
For a third-rank tensor $Y_{ijk}$ with symmetry in the last two indices, i.e.,
$Y_{ijk}=Y_{i(jk)}$, it can be decomposed into the following form \cite{OVEH97}
\begin{equation}\label{decom}
Y_{ijk}=\mbox{\textyen}_{ijk}+\frac{3}{5}\gamma_{i\langle j}L_{k\rangle}
-\frac{1}{5}\gamma_{i\langle j}Y_{k\rangle}+\frac{1}{3}\gamma_{jk}Y_i,
\end{equation}
where
\begin{equation}
L_i\equiv\gamma^{jk}Y_{jki}=\gamma^{jk}Y_{jik},\quad
Y_i\equiv\gamma^{jk}Y_{ijk},
\end{equation}
and \textyen$_{ijk}$ is the traceless part of $Y_{ijk}$, i.e.,
$\mbox{\textyen}_{ik}{}^{k}=\mbox{\textyen}^k{}_{ik}=\mbox{\textyen}^k{}_{ki}
=0$.
We can apply the decomposition (\ref{decom}) to the connection in
{\it every} time slice although a connection is not a tensor.
The is because any quantity can be decomposed like a tensor
as long as the quantity is only considered in the same coordinate,
without any coordinate transformations involved.
Thus the conformal connection can be decomposed as
\begin{equation}\label{Gdecom}
{\tilde\Gamma}^i{}_{jk}={\tilde F}^i{}_{jk}+\frac{3}{5}\delta^i{}_{\langle j}
{\tilde T}_{k\rangle}-\frac{1}{5}
\delta^i{}_{\langle j}{\tilde\Gamma}^{\bf g}_{k\rangle}
+\frac{1}{3}{\tilde\gamma}_{jk}{\tilde\Gamma}^i_{\bf g},
\end{equation}
where ${\tilde T}_i\equiv{\tilde\Gamma}^k{}_{ki}={\tilde\Gamma}^k{}_{ik}$,
and ${\tilde F}^i{}_{jk}$ is the traceless part of ${\tilde\Gamma}^i{}_{jk}$.
Here ${\tilde T}_i={\tilde\Gamma}^k{}_{ki}=\partial_i\ln\sqrt{\tilde\gamma}=0$
analytically, but ${\tilde T}_i$ could be nonzero numerically due to the
truncation error.
We can guarantee the vanishing of ${\tilde T}_i$ by subtracting the terms
having it from Eq.~(\ref{Gdecom}).
${\tilde\Gamma}^i_{\bf g}$ in Eq.~(\ref{Gdecom}) can be replaced with the
conformal connection function $\tilde\Gamma^i$ by adding the
$\Gamma$-constraint to it. Then the new connection becomes
\begin{align}
\tilde{\boldsymbol{\mit\Gamma}}{}^i{}_{jk}=&{\tilde\Gamma}^i{}_{jk}
-\frac{3}{5}\delta^i{}_{\langle j}{\tilde T}_{k\rangle}
-\frac{1}{5}\delta^i{}_{\langle j}{\mathcal G}_{k\rangle}+\frac{1}{3}
{\tilde\gamma}_{jk}{\mathcal G}^i\nonumber\\
=&{\tilde F}^i{}_{jk}-\frac{1}{5}\delta^i{}_{\langle j}{\tilde\Gamma}_{k\rangle}
+\frac{1}{3}{\tilde\gamma}_{jk}{\tilde\Gamma}^i.\label{newG}
\end{align}
We substitute $\tilde\Gamma^i{}_{jk}$ with the new one
$\tilde{\boldsymbol{\mit\Gamma}}{}^i{}_{jk}$ in all the calculations, including
the conformal covariant derivatives and the conformal Ricci tensor,
as a new modification.

Similarly, the spatial derivative of the conformal metric can also be
decomposed in every time slice as
\begin{equation}\label{gdecom}
\partial_i{\tilde\gamma}_{jk}=\overline{\partial_i{\tilde\gamma}}_{jk}
+\frac{3}{5}{\tilde\gamma}_{i\langle j}{\tilde\Gamma}^{\bf g}_{k\rangle}
-\frac{1}{5}{\tilde\gamma}_{i\langle j}T_{k\rangle}
+\frac{1}{3}{\tilde\gamma}_{jk}T_i,
\end{equation}
where $\overline{\partial_i{\tilde\gamma}}_{jk}$ is the traceless part of
$\partial_i{\tilde\gamma}_{jk}$.
We use the same trick on the conformal connection part like we did on
Eq.~(\ref{newG}) and obtain a new spatial derivative on ${\tilde\gamma}_{ij}$
\begin{equation}
{\rm d}^{\prime}_i{\tilde\gamma}_{jk}=\partial_i{\tilde\gamma}_{jk}
+\frac{3}{5}{\tilde\gamma}_{i\langle j}{\mathcal G}_{k\rangle}.\label{newpgp}
\end{equation}
Here we do not take action on eliminating the $T^i$ part in Eq.~(\ref{gdecom})
since its effect is negligible from the observation of our numerical experiment.

However, it turns out that this replacement of the spatial derivative of the
conformal metric seems give too much change on the field equation of
$\tilde\gamma_{ij}$ and causes in instability when
${\rm d}^{\prime}_i{\tilde\gamma}_{jk}$ is applied in Eq.~(\ref{lieg}).
This problem can be solved when the modification (\ref{modta}) in
Sec.~\ref{appmom} is applied in simulations, at least in single black hole
simulations.
Nevertheless, we modify Eq.~(\ref{newpgp}), by multiplying an adjustable
parameter to the substitution part, to have
\begin{equation}
{\rm d}_i{\tilde\gamma}_{jk}=\partial_i{\tilde\gamma}_{jk}
+\sigma{\tilde\gamma}_{i(j}{\mathcal G}_{k)}
-\frac{1}{5}{\tilde\gamma}_{jk}{\mathcal G}_i,\label{newpg}
\end{equation}
where the range of the parameter is usually chosen as $\sigma\in[1/5,4/5]$
in this work. And
\begin{equation}
\beta^i{\rm d}_i{\tilde\gamma}_{jk}=\beta^i\partial_i{\tilde\gamma}_{jk}
+\sigma\beta_{(j}{\mathcal G}_{k)}
-\frac{1}{5}{\tilde\gamma}_{jk}\beta^i{\mathcal G}_i.
\end{equation}
Thus the replacement of $\beta^i\partial_i{\tilde\gamma}_{jk}$ in
Eq.~(\ref{lieg}) with $\beta^i{\rm d}_i{\tilde\gamma}_{jk}$ is equivalent to
modifying Eq.~(\ref{dtg}) into
\begin{equation}\label{dtg2}
\frac{\rm d}{{\rm d}t}\tilde{\gamma}_{ij}=-2\alpha\tilde{A}_{ij}
+\sigma\beta_{(i}{\mathcal G}_{j)}
-\frac{1}{5}{\tilde\gamma}_{ij}\beta^k{\mathcal G}_k.
\end{equation}
We postpone the modification on the field equation of ${\tilde A}_{ij}$
with the decomposition of the spatial derivative of ${\tilde A}_{ij}$ until
Sec.~\ref{appmom}.
\subsection{Enforcement of the $\Gamma$-constraint}\label{Gammamod}
In the conventional adjustments, to enforce constraint
(\ref{Gofg}) all the undifferentiated $\tilde\Gamma^i$ in the
evolution equations are substituted with $\tilde\Gamma^i_{\bf g}$.
However, this adjustment give stability in single Kerr-Schild black hole
simulations only  in octant symmetry \cite{almb01}.
Instead of the conventional adjustment, one of the alternative adjustments
\cite{yhbs02} (dubbed as ``YBS'') is to add the $\Gamma$-constraint to
the evolution equation (\ref{dtGamma}) of ${\tilde\Gamma}^i$ by
\begin{equation}\label{dtGamma1}
\partial_t{\tilde\Gamma}^i=\mbox{rhs of }(\ref{dtGamma})-\frac{2}{3}(\xi+1)
{\mathcal G}^i\beta^k{}_{,k},
\end{equation}
where $\xi$ is usually chosen to be unity.
The YBS adjustment has been proven to be helpful in suppressing the
instability caused from some unstable modes \cite{yhbs02,SpeU07}.

Here we propose another adjustment to enhance the stability.
This adjustment is basically the following substitution in
Eq.~(\ref{dtGamma}):
\begin{equation}
\tilde\gamma^{ij}K_{,j}\rightarrow\kappa^{ij}{}_{,j}+K\tilde\Gamma^i,
\end{equation}
where $\kappa^{ij}\equiv\tilde\gamma^{ij}K$.
And this adjustment turns Eq.~(\ref{dtGamma}) into
\begin{align}
\partial_t\tilde{\Gamma}^i=&2\alpha(
{\tilde\Gamma}^i{}_{jk}\tilde A^{jk}-
\frac{2}{3}\kappa^{ij}{}_{,j}+6\tilde{A}^{ij}\phi_{,j})-
2\tilde{A}^{ij}\alpha_{,j}\nonumber\\
&+\beta^j\tilde{\Gamma}^i{}_{,j}-\tilde{\Gamma}^j\beta^i{}_{,j}
+\tilde{\gamma}^{jk}\beta^i{}_{,jk}+\frac{1}{3}\tilde{\gamma}^{ij}
\beta^k{}_{,jk}\nonumber\\
  &+\frac{2}{3}(\beta^k{}_{,k}-2\alpha K)\tilde{\Gamma}^i.\label{dtGamma2}
\end{align}
As we will show in Sec.~\ref{nr}, the performance of the code
with this adjustment is better than the one with the YBS adjustment,
but without the uncertainty of choosing the value of the parameter $\xi$.
Furthermore, the connection in Eq.~(\ref{dtGamma2}) can be replaced with the
new connection $\tilde{\boldsymbol{\mit\Gamma}}{}^i{}_{jk}$, and thus
Eq.~(\ref{dtGamma2}) is modified into
\begin{align}
\partial_t\tilde{\Gamma}^i=&2\alpha(
\tilde{\boldsymbol{\mit\Gamma}}^i{}_{jk}\tilde A^{jk}-
\frac{2}{3}\kappa^{ij}{}_{,j}+6\tilde{A}^{ij}\phi_{,j})-
2\tilde{A}^{ij}\alpha_{,j}\nonumber\\
 &+\beta^j\tilde{\Gamma}^i{}_{,j}-\tilde{\Gamma}^j\beta^i{}_{,j}
  +\tilde{\gamma}^{jk}\beta^i{}_{,jk}+\frac{1}{3}\tilde{\gamma}^{ij}
 \beta^k{}_{,jk} \nonumber\\
  &+\frac{2}{3}(\beta^k{}_{,k}-2\alpha K)\tilde{\Gamma}^i
  -(1+\xi)\Theta(\lambda^i)\lambda^i{\mathcal G}^i,\label{dtGamma3}
\end{align}
where the last term is newly added to control the stability via the linear term
of ${\tilde\Gamma}^i$, $\Theta$ is a step function, and $\lambda^i$ is as
follows:
\begin{equation}
\lambda^i=\frac{2}{3}(\beta^k{}_{,k}-2\alpha K)-\beta^{\hat i}{}_{,\hat i}
-\frac{2}{5}\alpha{\tilde A}_{\hat i}{}^{\hat i},
\end{equation}
where the index with hat, i.e., $\hat i$, means that no index summation happens
on this index. $\xi$ is chosen to be $1$ in all the cases in this work.
\subsection{Application of the momentum constraint}\label{appmom}
Yoneda \& Shinkai have studied the adjusted systems for the BSSN
formulation in \cite{ygsh02}.
It shown in the work that the adjusted BSSN system could be quite robust
with the following modification on the field equation (\ref{dtA}) of
${\tilde A}_{ij}$ :
\begin{equation}
\frac{\rm d}{{\rm d}t}\tilde{A}_{ij}=\mbox{rhs of }(\ref{dtA})
+\kappa_A\alpha{\tilde\nabla}_{(i}{\mathcal M}_{j)},
\end{equation}
where $\kappa_A$ is a constant. If $\kappa_A$ is set as positive, the
violations of the constraints are expected to be damped.
The robustness of this $\tilde A$-adjusted BSSN formulation has been
demonstrated in \cite{kksh08,ttys12}.
Here we would like to apply this type of modification to the BSSN formulation
in a slightly different manner.
In the momentum constraint, the covariant derivative of ${\tilde A}_{ij}$ can
be rewritten as
\begin{equation}
\tilde{\nabla}_j\tilde{A}_i{}^j=\tilde{A}_i{}^j{}_{,j}-
{\tilde\Gamma}^{kj}{}_i\tilde{A}_{kj}=
\tilde{A}_i{}^j{}_{,j}-\frac{1}{2}{\tilde\gamma}^{jk}\tilde{A}_{jk,i}.
\end{equation}
Thus the momentum constraint turns to be
\begin{equation}
{\mathcal M}_i=\tilde{A}_i{}^j{}_{,j}-
               \frac{1}{2}{\tilde\gamma}^{jk}\tilde{A}_{jk,i}+
               6\phi_{,j}\tilde{A}_i{}^j-\frac{2}{3}K_{,i}.\label{mom3}
\end{equation}
And the symmetric part of its spatially partial derivative is
\begin{align}
{\mathcal M}_{(i,j)}=&\tilde{A}_{(i}{}^k{}_{,j)k}+
               {\tilde\Gamma}^{k\ell}{}_{(i|}\tilde{A}_{k\ell,|j)}-
               \frac{1}{2}{\tilde\gamma}^{k\ell}\tilde{A}_{k\ell,ij}\nonumber\\
         &+6\phi_{,k(i}\tilde{A}_{j)}{}^k+6\phi_{,k}\tilde{A}_{(i}{}^k{}_{,j)}-
               \frac{2}{3}K_{,ij}.
\end{align}
Therefore, the modification on Eq.~(\ref{dtA}) in this work is
\begin{equation}\label{modta}
\frac{\rm d}{{\rm d}t}\tilde{A}_{ij}=\mbox{rhs of }(\ref{dtA})+hf(\alpha)
{\mathcal M}_{\langle i,j\rangle},
\end{equation}
where $h$ is the grid width and $f(\alpha)$ is a function of lapse and
usually chosen to be $1$ in single black hole simulations.
It shows in our numerical experiments that this modification is helpful in
suppressing the instability from the high-frequency unstable modes.

As in Sec.~\ref{dp}, the spatial derivative of ${\tilde A}_{ij}$ can be
decomposed in every time slice as
\begin{equation}\label{Adecom}
\partial_i{\tilde A}_{jk}=\overline{\partial_i{\tilde A}}_{jk}
+\frac{3}{5}{\tilde\gamma}_{i\langle j}P_{k\rangle}
-\frac{1}{5}{\tilde\gamma}_{i\langle j}Q_{k\rangle}
+\frac{1}{3}{\tilde\gamma}_{jk}Q_i,
\end{equation}
where $\overline{\partial_i{\tilde A}}_{jk}$ is the traceless part of
$\partial_i{\tilde A}_{jk}$ and
\begin{equation}
P_i={\tilde\gamma}^{jk}\partial_j{\tilde A}_{ki},\quad
Q_i={\tilde\gamma}^{jk}\partial_i{\tilde A}_{jk}.
\end{equation}
With the momentum constraint (\ref{mom3}) and the traceless extrinsic curvature
constraint (\ref{trA0}), we obtain a new spatial derivative on
${\tilde A}_{ij}$
\begin{equation}
{\rm d}_i{\tilde A}_{jk}=\partial_i{\tilde A}_{jk}
-\frac{3}{5}{\tilde\gamma}_{i\langle j}{\mathcal M}_{k\rangle}
-\frac{1}{10}{\tilde\gamma}_{i\langle j}{\mathcal A}_{k\rangle}
-\frac{1}{3}{\tilde\gamma}_{jk}{\mathcal A}_i,\label{newdA}
\end{equation}
where ${\mathcal A}_i$ is the spatially secondary constraint of
Eq.~(\ref{trA0})
\begin{equation}
{\mathcal A}_i=\partial_i{\mathcal T}={\tilde\gamma}^{jk}{\tilde A}_{jk,i}
+{\tilde A}_{jk}{\tilde\gamma}^{jk}{}_{,i}=Q_i
-2{\tilde A}_{jk}\tilde{\boldsymbol{\mit\Gamma}}{}^{jk}{}_i=0.
\end{equation}
With the new spatial derivative (\ref{newdA}), we can modify the field equation
of ${\tilde A}_{ij}$ further from Eq.~(\ref{modta}) to
\begin{align}
\frac{\rm d}{{\rm d}t}\tilde{A}_{ij}=\mbox{rhs of }(\ref{dtA})&+hf(\alpha)
{\mathcal M}_{\langle i,j\rangle}-\frac{3}{5}\beta_{\langle i}
{\mathcal M}_{j\rangle}\nonumber\\
&-\frac{1}{10}\beta_{\langle i}{\mathcal A}_{j\rangle}
-\frac{1}{3}{\tilde\gamma}_{ij}\beta^k{\mathcal A}_k.\label{modta2}
\end{align}
\section{Initial Data}
\subsection{Single black hole in Kerr-Schild coordinates}\label{idfiksbh}
The ingoing Kerr-Schild form of the Kerr metric is given by
\cite{mrhs99,chas92}
\begin{equation}\label{ksf1}
{\rm d}s^2=(\eta_{\mu\nu} + 2H\ell_\mu\ell_\nu){\rm d}x^\mu{\rm d}x^\nu,
\end{equation}
where $\eta_{\mu\nu}={\rm diag}(-1,1,1,1)$ is the Minkowski metric in
Cartesian coordinates, and $H$ a scalar function.
The vector $\ell_\mu$ is null both with respect to $\eta_{\mu\nu}$ and
$g_{\mu\nu}$,
\begin{equation}
\eta^{\mu\nu}\ell_\mu\ell_\nu=g^{\mu\nu}\ell_\mu\ell_\nu=0,
\end{equation}
and we have $\ell^2_t=\ell^k\ell_k$.
The general Kerr-Schild black hole metric has
\begin{equation}
H=\frac{Mr}{r^2+a^2\cos^2\theta} \label{3}
\end{equation}
and
\begin{equation}
\ell_{\mu}=\left(1,\frac{rx+ay}{r^2+a^2},\frac{ry-ax}{r^2+a^2},\frac{z}{r}
\right).  \label{4}
\end{equation}
Here $M$ is the mass of the Kerr black hole, $a=J/M$ is the specific angular
momentum of the black hole, and $r$ and $\theta$ are auxiliary spheroidal
coordinates defined in terms of the Cartesian coordinates by
\begin{equation}
\frac{x^2+y^2}{r^2+a^2}+\frac{z^2}{r^2}=1\label{5}
\end{equation}
and $z=r\cos\theta$. The event horizon of the black hole is located at
\begin{equation}
r_{\rm eh} = M+\sqrt{M^2-a^2}.\label{reh}
\end{equation}

Comparing (\ref{ksf1}) with the ADM metric (\ref{adm_metric}) one
identifies the lapse function $\alpha$, shift vector $\beta_i$ and the
spatial 3-metric $\gamma_{ij}$ as
\begin{align} \label{ksf2}
              \alpha &=\frac{1}{\sqrt{1+2H}}, \\
             \beta_i &=2H\ell_i,\\
         \gamma_{ij} &=\eta_{ij} + 2H\ell_i\ell_j.
\end{align}
We can see here that these variables all extend smoothly through the horizon
and their gradients near the horizon are well-behaved.
Given these metric quantities, the extrinsic curvature $K_{ij}$
can be computed from (\ref{gdot1})
\begin{align}
K_{ij}&=2\alpha H\ell^k[\ell_i\ell_jH_{,k}+2H\ell_{(i}\ell_{j),k}]\nonumber\\
     &+2\alpha[\ell_{(i}H_{,j)}+H\ell_{(i,j)}],\\
    K&=2\alpha^3(1+H)\ell^k H_{,k}+2\alpha H\ell^k{}_{,k}.
\end{align}

In the static case $a=0$, the above expressions reduce to the
Schwarzschild expressions in ingoing Eddington-Finkelstein form \cite{edda58}
\begin{align}
      H &=\frac{M}{r},\quad \ell_\mu=(1,\frac{x_k}{r}),\nonumber\\
      K_{ij}&=\frac{2M}{ r^4(1+2M/r)^{1/2}}\left[r^2\eta_{ij}
       -(2+\frac{M}{r})x_ix_j\right],\\
      K&= \frac{2M}{r^2(1+2M/r)^{3/2}}(1+3\frac{M}{r}),\nonumber
\end{align}
where $M$ is the total mass-energy and $r^2 = x^2 + y^2 + z^2$.
\subsection{Binary black hole with punctures}
\begin{table*}[t]
\caption{Input parameters and modifications used for selected evolutions.
For each evolution we list the symmetry used,
the method for constraints $\mathcal D$ \& $\mathcal T$ enhancement,
the modifications on the field equations of ${\tilde\Gamma}^i$ and
${\tilde\gamma}_{ij}$,
the substitution of connection,
the modification on the field equations of ${\tilde A}_{ij}$,
the time when instability shows.}
\begin{ruledtabular}
\begin{tabular}{lccccccc}
Case
&Symmetry
&${\tilde\gamma}_{zz}$+${\tilde A}_{yy}$
&${\widetilde\Gamma}^i$
&${\tilde\gamma}_{ij}$
&$\tilde{\boldsymbol{\mit\Gamma}}{}^i{}_{jk}$
&${\tilde A}_{ij}$
&Instability\\
Eqn&
&(\ref{gzz})+(\ref{Ayy})
&(\ref{dtGamma3})
&(\ref{dtg2})
&(\ref{newG})
&(\ref{modta2})
&appears at\\
 \hline
STD&Equ&(\ref{Agave})&${\tilde\Gamma}^i_{\bf g}$&&&&$\sim 250M$\\
YBS&Equ&$\surd$&(\ref{dtGamma1})&&&&$\sim 650M$\\
E1 &Equ&$\surd$&(\ref{dtGamma2})&&&&$\sim 550M$\\
E2 &Equ&$\surd$&(\ref{dtGamma2})&$\sigma$=2/5&&&$\sim 750M$\\
E3 &Equ&$\surd$&$\surd$&$\sigma$=2/5&$\surd$&&$\sim 850M$\\
E4 &Equ&$\surd$&$\surd$&$\sigma$=2/5&$\surd$&(\ref{modta})&$\sim 950M$\\
E5 &Equ&$\surd$&$\surd$&$\sigma$=3/5&$\surd$&(\ref{modta})&None\\
E6-lo&Equ&$\surd$&$\surd$&$\sigma$=3/5&$\surd$&$\surd$&None\\
E6&Equ&$\surd$&$\surd$&$\sigma$=3/5&$\surd$&$\surd$&None\\
E6-hi&Equ&$\surd$&$\surd$&$\sigma$=4/5&$\surd$&$\surd$&None\\
N7&None&$\surd$&$\surd$&$\sigma$=3/5&$\surd$&$\surd$&None
\end{tabular}
\end{ruledtabular}
\label{tab1}
\end{table*}
We use the quasi-circular binary black hole puncture data to test our
modifications.
The momentum parameter for the quasi-circular orbit is set to be the value
given by \cite{tichy04} based on the helical Killing vector conditions.
In the following, we review the puncture scheme and describe how we construct
initial data by the multi-domain spectral method extended from the LORENE
library \cite{lorene,loreneweb}.

To determine a three-geometry subject to the constraint equations (\ref{ham1})
and (\ref{mom1}), the conformal decomposition plays an important role (see
\cite{IDlecture} for instance).
Lichnerowicz \cite{CTT1} proposed a conformal decomposition of three-metric
$\gamma_{ij}=\psi^{4}\tilde{\gamma}_{ij}$, as Eq.~(\ref{confgij}) with
$\psi=e^\phi$, and later York \cite{CTT2} used the transverse-traceless
decomposition of the conformal extrinsic curvature
$\hat{A}^{ij}\equiv\psi^{10}A^{ij}=\hat{A}_{TT}^{ij}+\tilde{\mathbb{L}}W^{ij}$.
The particular conformal scaling makes the identity
$\nabla_{i}A^{ij}=\psi^{-10}\tilde{\nabla}_{i}\hat{A}^{ij}$ hold.
In this decomposition, $\hat{A}_{TT}^{ij}$ is transverse (i.e.,
$\tilde{\nabla}_{i}\hat{A}_{TT}^{ij}=0$) and can be assumed to be zero to have
a purely longitudinal $\hat{A}^{ij}=\tilde{\mathbb{L}}W^{ij}$.
In this approach, named as conformal transverse-traceless method (CTT), and in
the assumption of conformal flatness, $\tilde{\gamma}_{ij}=\eta_{ij}$,
maximal slicing, $K=0$, and $\hat A^{ij}_{TT}=0$, the vacuum constraint
equation would then be expressed as
\begin{align}
\tilde{\nabla}^{2}\psi &=-\frac{1}{8}\psi^{-7}\hat{A}_{ij}\hat{A}^{ij},\\
\tilde{\nabla}_{j}\hat A^{ij} &=0.
\end{align}
The momentum constraint is linear in this case. For a $N$-black hole system,
it allows the Bowen-York solution $\hat{A}^{ij}=\sum_{a=1}^N\hat{A}_a^{ij}$,
where the conformal extrinsic curvature for each black hole is \cite{BJYJ80}
\begin{align}
\hat{A}_a^{ij}\equiv\psi^{10}A_a^{ij}&=\frac{3}{4r^2}[P^{(i}n^{j)}-
2(\gamma^{ij}-n^i n^j)P_k n^k] \nonumber \\
&+\frac{3}{2r^3}n^{(i}\epsilon^{j)k\ell}S_k n_\ell,
\end{align}
with $n^i$ the spatial unit vector pointing away from the puncture,
and $P^i$ and $S^i$ its linear and intrinsic angular momentum respectively.
The conformal factor then can be numerically solved from the Hamiltonian
constraint.

In the puncture method, one separates out the singular part
\begin{equation}\label{singular}
\psi_s=1+\sum_{a=1}^N\frac{m_a}{2r_a},
\end{equation}
from the conformal factor $\psi$, where $m_{a}$ and $r_{a}$ are the mass
parameter and coordinate distance from each puncture respectively.
It is the superposition of Schwarzschild solution in the isotropic coordinate
and satisfies the Laplace equation \cite{BDLR63}.
Therefore the desired regular part $u\equiv\psi-\psi_s$ is determined by
\begin{equation}
\tilde{\nabla}^2 u=-\frac{1}{8}\hat{A}^{ij}\hat{A}_{ij}(\psi_s+u)^{-7}.
\label{puncture_eq}
\end{equation}
The existence and uniqueness of the solution has been discussed in
\cite{brandt97}.

Our multi-puncture initial data solver is motivated from \cite{GEGB02} for the
excised binary black hole initial data. We cover on each black hole a
spherical multi-shell domain, and split $u=\sum_a u_a$ (the index $a$ runs over
the number of the black holes) as well as the puncture equation
(\ref{puncture_eq}) into
\begin{equation}
\tilde{\nabla}^2 u_a=-\frac{1}{8}\hat{A}_a^{ij}\hat{A}_{ij}(\psi_s+u)^{-7}.
\end{equation}
Only one of the extrinsic curvature tensor $\hat{A}_{ij}$ split in the above
equation.
Thus it is expected that the source term in its RHS has large contribution only
near each hole, and the use of spherical polar coordinates is adequate for
solving the equation near the punctures.
The $u_a$ vanished at the asymptotically flat physical outer boundary on the
outermost, compactified shell.
And $\partial_r u_a$($r=0$)$=0$ is also ensured at the punctures as the inner
boundary condition. These equations for each hole are then solved iteratively
with the Poisson solver in the LORENE library until each successive difference
of $\delta u_a$ is small, typically $10^{-11}$. For the binary case, our
initial data is consistent with the earlier work \cite{two_puncture}.
The convergence of our method was presented in \cite{3BH}.
\section{Numerical Implementation}\label{ni}
In this work, we discretize the evolution equations using a fourth-order
Runge-Kutta scheme.
We use fourth-order centered differencing everywhere except for the advection
terms on the shift.
For these terms, a fourth-order upwind scheme is used along the shift direction.

For the lapse gauge condition, we consider the ``1+log'' slicing
\cite{bmss95,aaea99,amea00} which is basically a modification of the
Bona-Mass\'o slicing.
The lapse condition used in the single Kerr-Schild black hole evolutions is
\begin{equation} \label{1+log}
\partial_t\alpha=D_i\beta^i-\alpha K.
\end{equation}
For binary black hole simulations, the moving puncture technique is adopted.
And the lapse condition used here is the standard one in numerical relativity
community for binary black hole simulation as
\begin{equation}
\partial_t\alpha=\beta^i\alpha_{,i}-2\alpha K.
\end{equation}

Many driver gauge conditions (e.g., the $\Gamma$-driver) for the shift vector
\cite{balakrishna96,meter06} are currently the main type of gauge conditions
used in the punctured black hole calculations.
In this work, we will only focus on these types of gauge conditions.
The hyperbolic type $\Gamma$-driver condition used for the single black hole
simulations is
\begin{align}
\partial_t\beta^i&=\frac{3}{4}B^i,\\
\partial_t B^i&=\alpha^2\partial_t\tilde{\Gamma}^i-\eta B^i,
\end{align}
where $\eta$ is the parameters to be chosen.
In the single Kerr-Schild black hole evolutions, $\eta=5$.
For the binary black hole simulations, we use
\begin{align}
\partial_t\beta^i&=\frac{3}{4}B^i+\beta^j\beta^i{}_{,j},\\
\partial_t B^i&=\partial_t\tilde{\Gamma}^i-2 B^i+\beta^jB^i{}_{,j}
-\beta^j \tilde{\Gamma}^i{}_{,j}.
\end{align}

On the outer boundaries of the numerical grid we impose a radiative boundary
condition that is imposed on the difference between a given variable and its
analytic value $f-f_{\rm analytic}=u(r-t)/r$ where $u$ is an outgoing wave
function.
We apply this condition to all fields except $\tilde{\Gamma}^i$ which we leave
fixed to their analytic values at the boundary in the single Kerr-Schild black
hole evolutions for stability.
We return to apply the radiative boundary condition on $\tilde{\Gamma}^i$ in
the punctured binary black hole evolutions.

The excision technique is applied in the single Kerr-Schild black hole
evolutions since the singularities in the Kerr-Schild coordinates are physical
ones for which the puncture method is not applicable.
(Usually the puncture method is applied to the isotropic-like coordinates in
which the singularities are coordinate ones.)
For (spherical) excision regions, we adopt the recipe suggested by
\cite{almb01,yhbs02} to copy the time derivative of every field at the
boundary from its neighboring grid-point.
The details can be found in \cite{yhbs02}.
\section{Numerical Result}\label{nr}
\subsection{Single black hole}
\begin{figure}[thbp]
\begin{tabular}{c}
\includegraphics[width=\columnwidth]{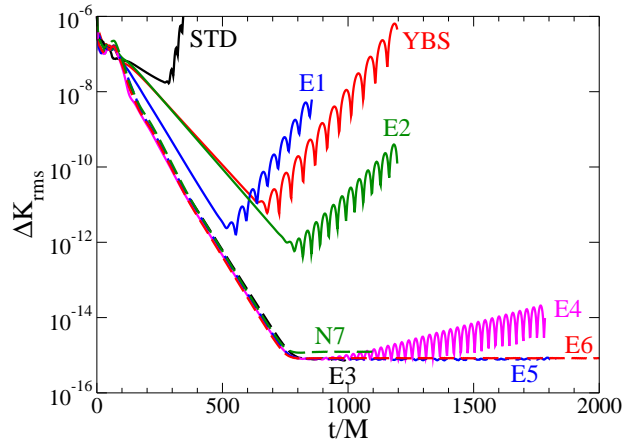}
\end{tabular}
\caption{The root mean square (rms) of the change in the trace of extrinsic curvature between
consecutive time steps as functions of time for the cases listed in Table
\ref{tab1}.
The usual BSSN formulation is used in Case STD. The modifications suggested
in \cite{yhbs02} are used in Case YBS.
The modifications suggested in this work are activated one by one from Case
E1 to Case E6.
Case E5 has the same modifications with Case E4 except with a different value
of $\sigma$ used in Eq.~(\ref{dtg2}).
Case N7 has the same modifications with Case E6 except with none symmetry
for the computational domain.
There is no instability shown in Cases E5, E6, and N7 throughout these runs.}
\label{fig1}
\end{figure}
\begin{figure*}[thbp]
\begin{tabular}{rl}
\includegraphics[width=\columnwidth]{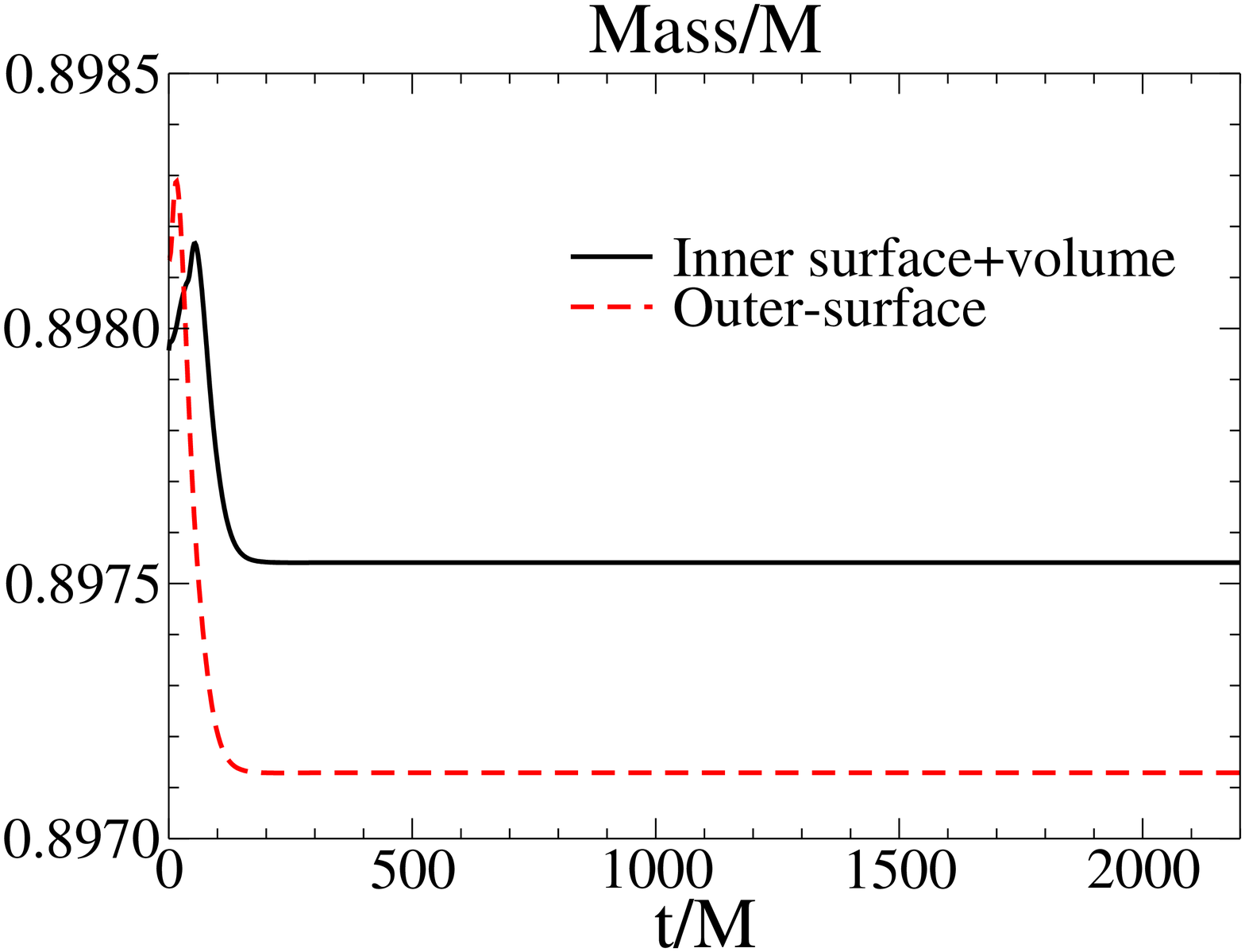}&
\includegraphics[width=\columnwidth]{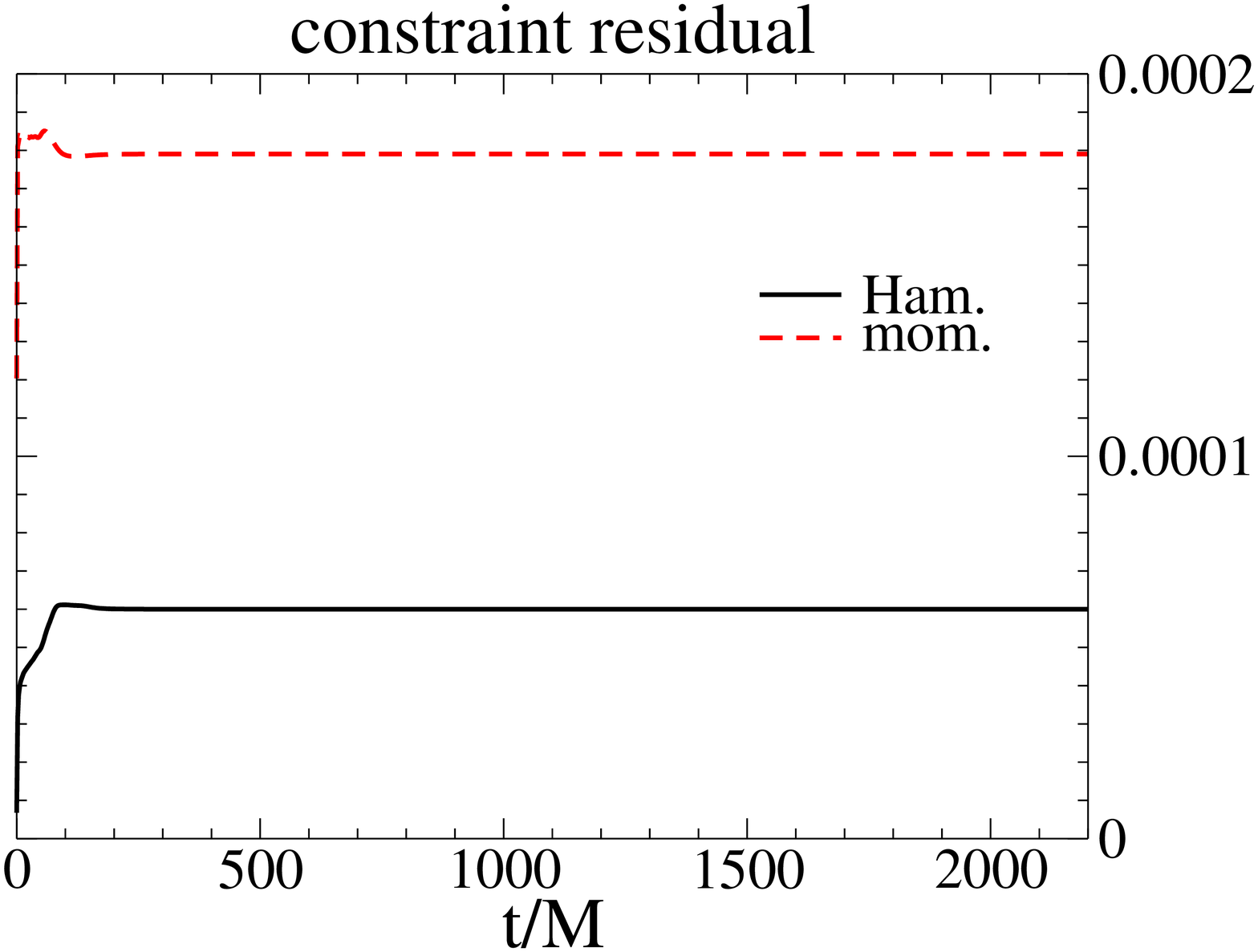}\\
\includegraphics[width=\columnwidth]{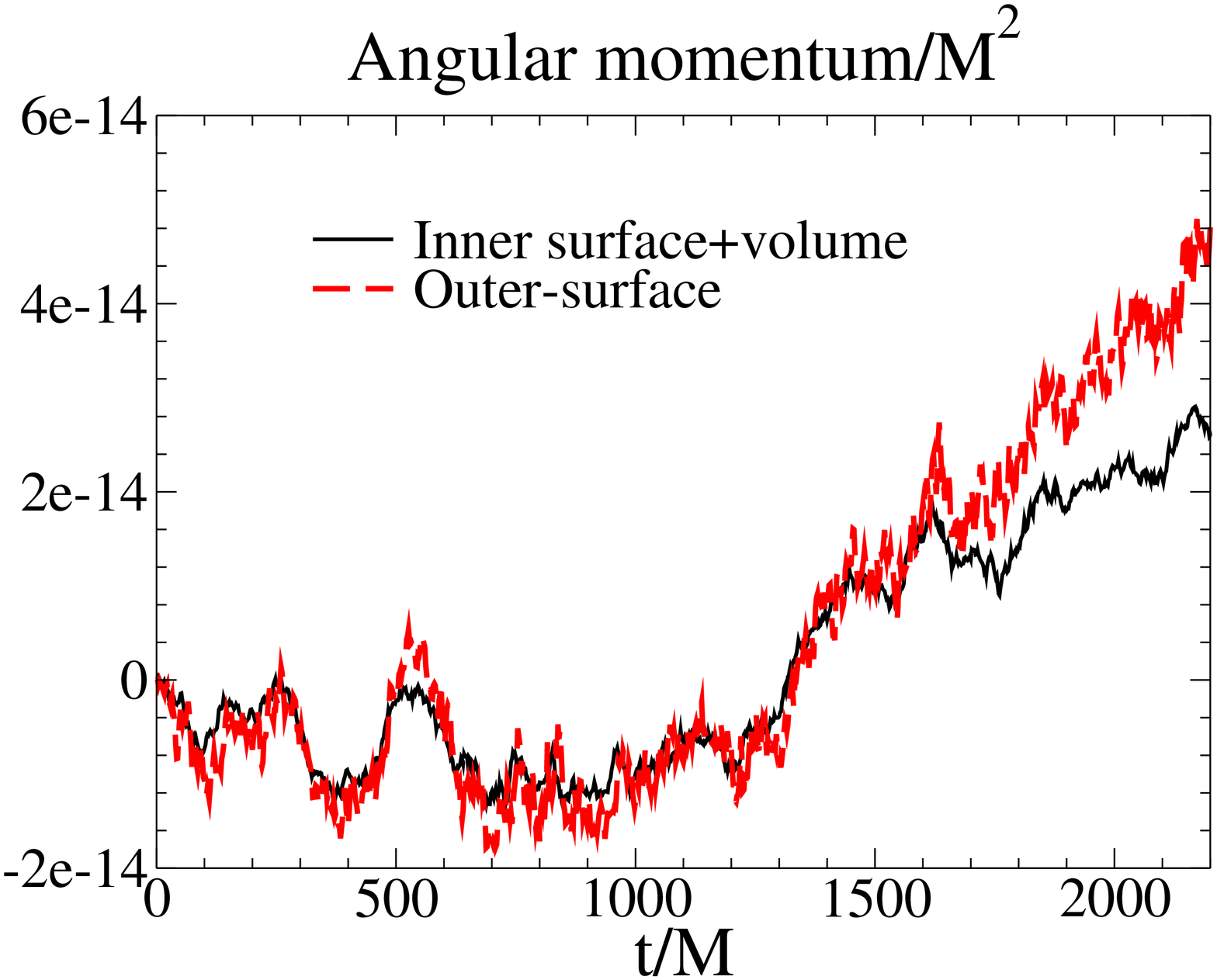}&
\includegraphics[width=\columnwidth]{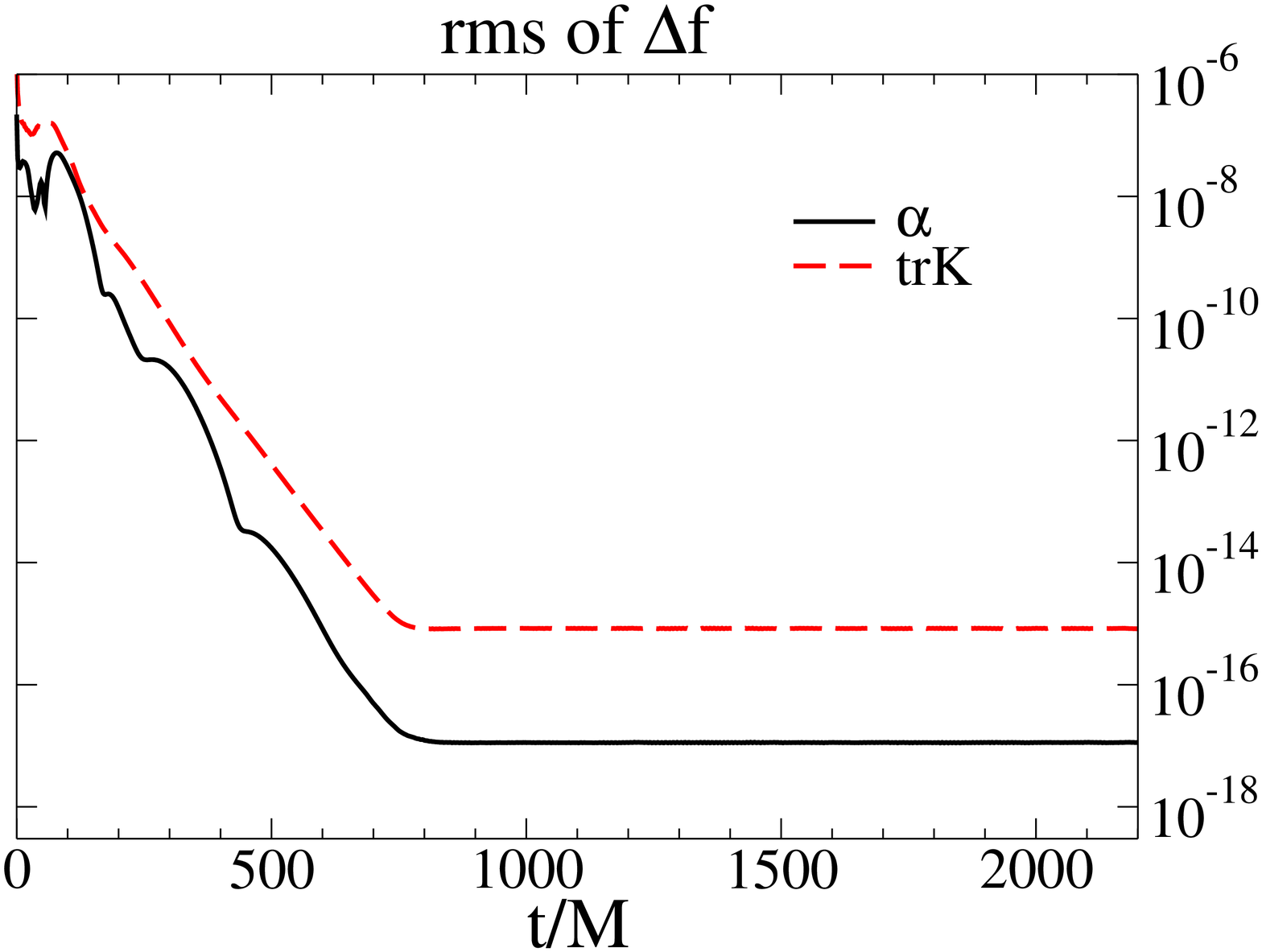}
\end{tabular}
\caption{The monitored quantities as functions of time for Case E6.
The upper-left panel compares different integrals for the ADM mass.
The lower-left panel compares different integrals for the angular momentum.
The upper-right panel shows the L2 norms of the Hamiltonian constraint
$\mathcal H$ and the momentum constraint ${\mathcal M}^x$.
The lower-right panel shows a log plot of the rms of the changes
in the lapse and the trace of extrinsic curvature between consecutive time
steps.}
\label{fig2}
\end{figure*}
\begin{figure}[thbp]
\begin{tabular}{c}
\includegraphics[width=\columnwidth]{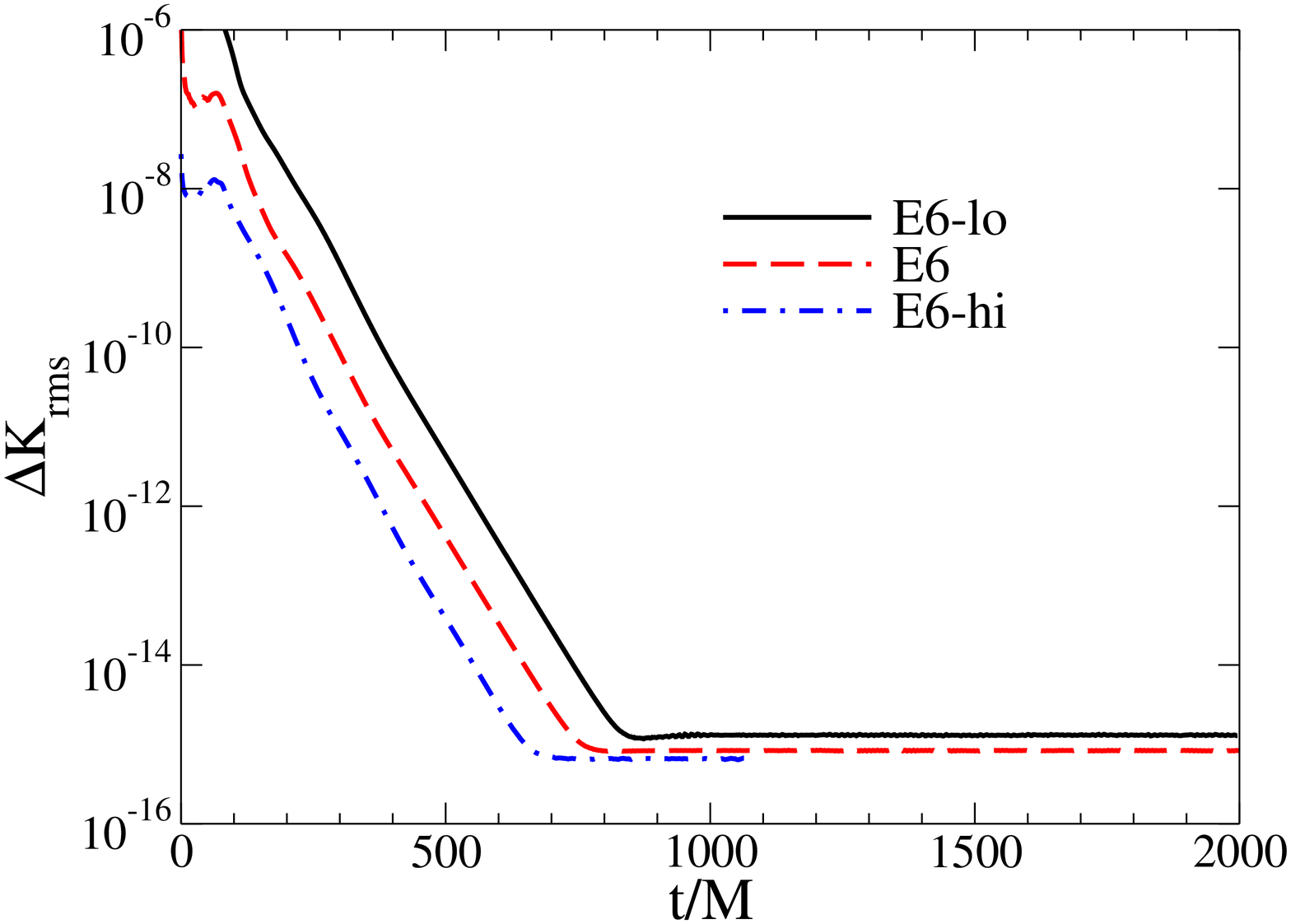}
\end{tabular}
\caption{The rms of the change in the trace of extrinsic curvature
between consecutive time steps as functions of time with different resolutions.
}
\label{fig3}
\end{figure}
Our simulations for single Kerr-Schild black holes are summarized in Table
\ref{tab1}.
For most of these simulations we use computational domains of size
$-12M<x$,$y<12M$ and $0<z<12M$ for equatorial symmetry,
and $-12M<x$,$y$,$z<12M$ for no symmetry, with a grid spacing of
$\Delta x=0.2M$.
In order to analyze the effect of resolution we performed the two Cases E6-lo
and E6-hi on the same domain and used a resolution of $\Delta x=0.4M$ for E6-lo
and $\Delta x=0.1M$ for E6-hi.
We always use a Courant factor of $1/4$ so that $\Delta t=\Delta x/4$.
We excise spheres of radius $r=1.6M$ inside the event horizons ($r=2M$) 
in these simulations.

A simulation can be judged as being stable if changes in all dynamical
variables drop to round-off error (of about $10^{-16}$ for double precision),
and remain at that level as the simulation goes.
Reaching round-off implies that the numerical solution has settled
down to the equilibrium solution of the finite-difference equations
(as opposed to the equilibrium solution of the differential equations,
which is provided as initial data).  In all our stable runs, besides
monitoring the global quantities consisting of the ADM mass (\ref{surfmass},
\ref{volmass}), the angular momentum $J_z$ (\ref{surfang},\ref{volang}),
the $L_2$ norms of the Hamiltonian constraint
$\mathcal H$, the momentum constraint ${\mathcal M}^x$, and the 
$\Gamma$-constraint ${\mathcal G}^x$ violations, we also monitor the changes in
the representative variables, i.e., $\phi$, $K$, $\tilde{\gamma}_{xx}$,
$\tilde{A}_{xx}$, ${\tilde\Gamma}^x$, $\alpha$, and $\beta^x$, to see if they
can reach and remain at the level of round-off error. 
Due to the existence of singularity inside the excision area,
we only consider the domain outside the excision (including the part inside the
horizon) for the estimate of the error norm and the changes.

In Fig.~\ref{fig1} we summarize the listed cases with the medium resolution
in Table \ref{tab1}.
As shown in \cite{yhbs02,almb01}, in Case STD the evolution with the usual
standard recipes applied to the BSSN formulation becomes unstable quickly.
The root mean square (rms) of the changes in $K$ between consecutive time steps,
$\Delta K_{\rm rms}$, in this case drops exponentially until $t\sim 250M$,
then at later times it increases exponentially.
This exponentially growing mode can be extrapolated back to about round-off
error at $t=0$, indicating that the mode is triggered by round-off error in the
initial data.
In Case YBS the recipes suggested in \cite{yhbs02}, including mainly
Eqs.~(\ref{gzz},\ref{Ayy}), and Eq.~(\ref{dtGamma1}), are applied to the
BSSN formulation.
Undoubtedly, the evolution in Case YBS is more stable than the one in Case
STD.
We can see it from that the $\Delta K_{\rm rms}$ in this case drops
exponentially until $t\sim 650M$ before it turns to increase exponentially.
In fact, the recipes used in Case YBS have been shown in \cite{yhbs02} to be
able to stabilize an evolution until its $\Delta K_{\rm rms}$ reaches round-off
error.
However, a lower grid resolution and the second-order finite differencing
methods both spatially and temporally are used in those numerical experiments.
The instability shown in Case YBS indicates that higher-frequency unstable
modes triggered with higher grid resolution and/or higher-order finite
differencing methods need to be tamed with further modifications to the BSSN
formulation.

We test our new modifications from Case E1 to Case E6.
In Case E1, the field Eq.~(\ref{dtGamma2}) for ${\tilde\Gamma}^i$, instead of
Eq.~(\ref{dtGamma1}) in Case YBS, is employed in the evolution.
As described in Sec.~\ref{Gammamod}, this modification is to enhance the
effect of the linear terms in the field equation of ${\tilde\Gamma}^i$ on
convergence and stability.
We can see from the plot that the evolution in Case E1 converges faster than
the one in Case YBS, and thus encounters the growth of instability from
round-off error earlier ($t\sim 550M$).
Besides the modifications employed in Case E1, we add the modified field
Eq.~(\ref{dtg2}) of ${\tilde\gamma}_{ij}$ in Case E2 with the parameter choice
$\sigma=2/5$.
In our numerical experience, this modification is a critical one among the new
modifications on the stability of an evolution.
The choice of the parameter $\sigma$ could also affect the behavior of
convergence.
From Fig.~\ref{fig1} we can see that the evolution in Case E2 does not show
any instability until $t\sim 750M$.
In Case E3, the connection ${\tilde\Gamma}^i{}_{jk}$ is substituted by the new
one, $\tilde{\boldsymbol{\mit\Gamma}}{}^i{}_{jk}$, via the application of
Eq.~(\ref{newG}).
The purpose of the modification is to substitute ${\tilde\Gamma}^i_{\bf g}$,
a decomposition part of ${\tilde\Gamma}^i{}_{jk}$, in ${\tilde\Gamma}^i{}_{jk}$
with the independent conformal connection ${\tilde\Gamma}^i$, as well as
enhancing the secondary constraint ${\tilde\Gamma}^j{}_{ij}\simeq 0$.
We can see in Fig.~\ref{fig1} that the combination of the modifications used
in Case E3 enhances quite a lot the stability of an evolution.
The evolution does not show any instability until $t\sim 850M$ and the
$\Delta K_{\rm rms}$ reaches almost round-off error before the unstable mode
appears.

However, we found that the modifications used in Case E3 are sensitive
to the grid resolution.
This leads us to employ the modification (\ref{modta}) in Cases E4 and E5.
Eq.~(\ref{modta}) is adopted from \cite{ygsh02} with some deformation.
This modification offers a dissipation effect on ${\tilde A}_{ij}$ which
fits into our requirement for stability.
With this modification, the evolution in Case E4 does not show instability
until $t\sim 950M$ with $\Delta K_{\rm rms}$ near round-off error.
We push the stability in Case E5 by setting the parameter used in
Eq.~(\ref{dtg2}) to be $\sigma=3/5$.
Then the $\Delta K_{\rm rms}$ and all the other changes drop
exponentially until they reach round-off error in Case E5.
In Case E6, the modification (\ref{modta2}) instead of Eq.~(\ref{modta}) is
employed.
This modification gives a little faster convergence.
The settings in Case N7 are the same as in Case E6 except there is not any
grid symmetry in N7.
This case is used to test the effect of the grid symmetry on the stability of
an evolution with the modifications.
It shows in Case N7 no sign of instability with the relaxation of symmetry.

Now we would like to look closer at the stable case, i.e., Case E6.
The results for Case E6 are presented in Fig.~\ref{fig2}.
The upper-left panel shows two different integrations of the ADM masses.
The (red) dashed line is computed from a surface integral (\ref{surfmass}) at
large separation, while the solid line is computed from a volume integral
(\ref{volmass}) plus a surface integral over a small sphere enclosing the
black hole singularity.
We choose a radius of $R_1 = 2M$ for the inner surface and $R_2 = 11M$ for the
outer surface. For $R_2 \rightarrow \infty$ the two mass integrals should agree
and should yield the analytic value $M$ of the initial data.
Our two mass integrals agree to within about 0.05\%.
The lower-left panel in Fig.~\ref{fig2} shows surface and volume integrations
of the angular momentum, similar to the mass integrations explained above.
The (red) dashed line is computed from the outer surface integral
(\ref{surfang}); the solid line is computed from a combination of volume
integral (\ref{volang}) and inner surface integral.
For both integrations the angular momentum is very close to zero,
as it is supposed to be.
These results indicate that the global quantities are consistent with the
expected values in single BH and are not sensitive to the existence of the
excision (as long as inside the BH horizon).

The upper-right panel shows the L2 norms of the Hamiltonian constraint
${\mathcal H}$ (solid line) and the momentum constraint ${\mathcal M}^x$
(red-dashed line).
The lower-right panel shows a log plot of the rms of the
changes in the lapse $\alpha$ (solid line) and the trace of extrinsic
curvature $K$ (red dashed line) between consecutive time steps.
The changes in $\alpha$ and $K$ both decrease exponentially until they reach
round-off error at about $t\sim 800M$.
We then continue to run the numerical simulation for this case till the time
being over $2000M$ and there is not any sign of instability.

In Fig.~\ref{fig3}, we compare the results of Cases E6's using the same
modifications but with different resolutions to test the convergence of the
formulations.
From the plot we can see that in all the three cases the $\Delta K_{\rm rms}$'s
reach the round-off errors.
The cases for the low and the medium resolutions are extended over $2000M$.
The case for the high resolution is terminated at $t\approx 1100M$ due to its
time-consuming computation.
In Case E6-hi we use $\sigma=4/5$, instead of $\sigma=3/5$ in Cases E6-lo and
E6, to enhance the stability against higher frequency noise.
The results of these three cases indicate a nice convergence with these
modifications.
\subsection{Binary black hole with punctures}
\begin{figure*}[ht]
\begin{tabular}{cc}
\includegraphics[width=0.5\textwidth]{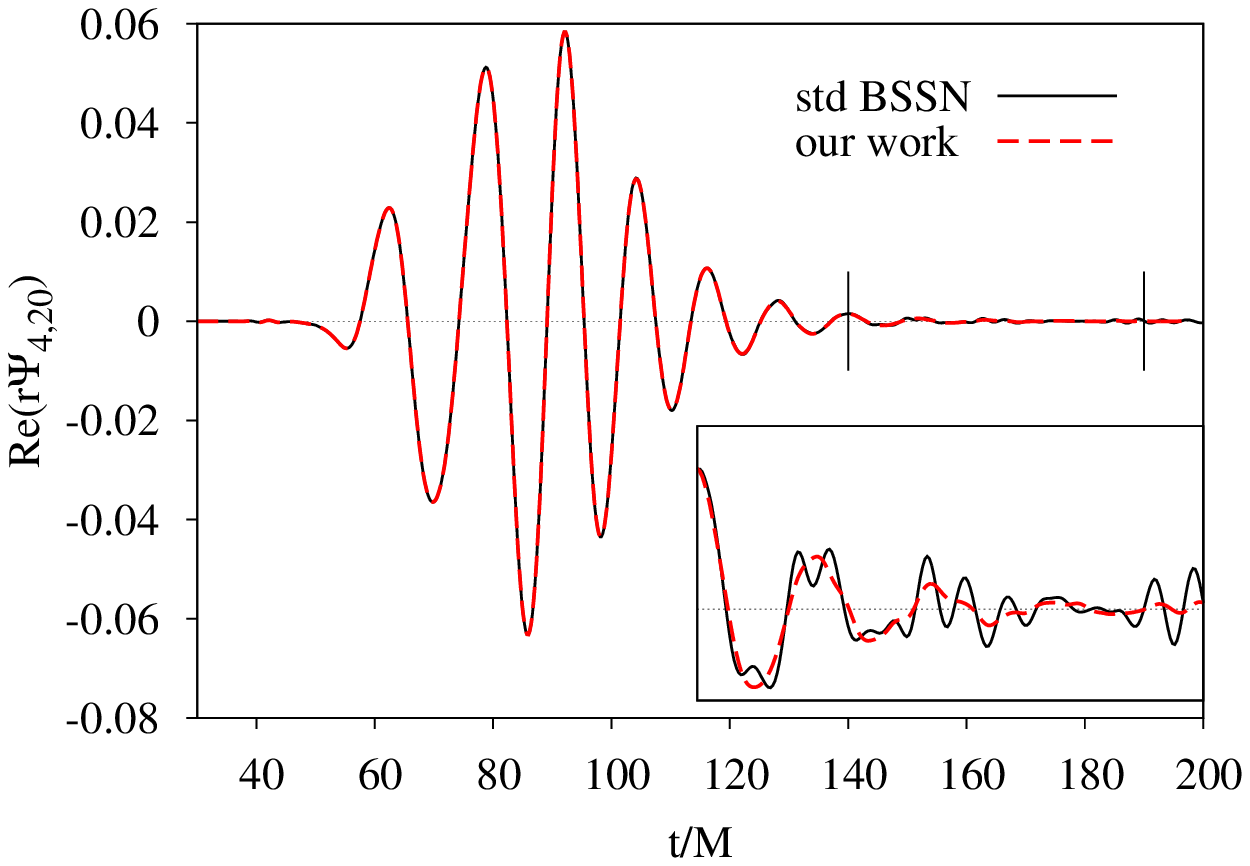}&
\includegraphics[width=0.5\textwidth]{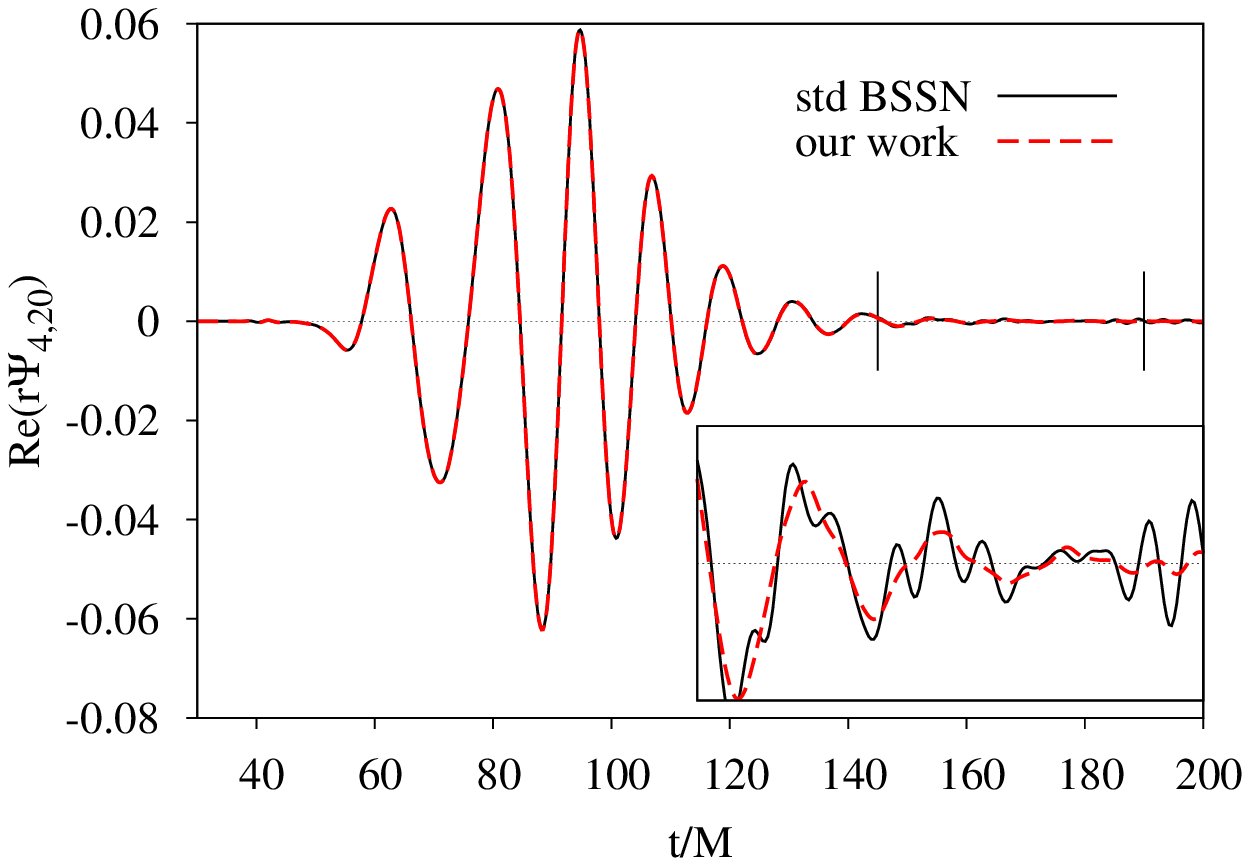}\\
\end{tabular}
\caption{Comparison of $\ell=2,m=2$ mode of $\Psi_4$ for the usual BSSN
formulation and the one with our new modification.
The outermost boundary locates at $517.12$. And the detector locates at $R=50$.
The (red) dashed line corresponds to the result with our new modification and
the solid line corresponds to the result with the usual BSSN formulation.
The results from the two formulations are quite consistent.
But for the late ringdown phase, shown in the enlarged subplots,
our new modification gives a much smoother result.}
\label{fig4}
\end{figure*}
We use our numerical code AMSS-NCKU, which has been employed in several
previous studies \cite{czyy08,cao10,cao12}. With this code the
standard moving box style mesh refinement is implemented. We used 11
mesh levels in all, where 8 levels are fixed and 3 levels are
movable. For fixed levels we used one box with grids
$144\times144\times72$ where we have taken the advantage of
equatorial symmetry of the system. The outermost physical boundary
is set to 517.12. For movable levels, two boxes are used. And every
box has grids $72\times72\times36$. In time direction, the
Berger-Oliger numerical scheme is adopted for levels higher than
$4$. Although our code can adopt the Kreiss-Oliger numerical
dissipation, we {\it disable} it in order to check the accuracy and
stability of our new modification. We set $\sigma=0.1$ in
Eq.~(\ref{dtg2}) and $f(\alpha)=h\alpha$ in Eq.~(\ref{modta}).

In this subsection both the usual BSSN formalism and our modification are
tested and compared with each other, without the Kreiss-Oliger numerical
dissipation for both formalisms.
Several typical configurations of binary black hole listed in the QC sequence
of \cite{baker02} are tested.
We find that, for QC1 and QC2, both formulations give stable and consistent
results for whole inspiral and merger phases.
However, for QC3 and QC6, both formulations crash at almost the same time
moment.
We suspect that the failure mainly comes from the numerical error being too
much.
Compared with the tests in the previous subsection,
there is one more complicated numerical issue, that is, the numerical noise
introduced by mesh refinement.
And this type of noise causes the numerical instability dominantly
when the mesh refinement technique is employed during the simulations.
Compared with the numerical noise by the discretization of the field equations,
the numerical noise induced by mesh refinement is much larger  such that the
modification is not able to suppress it in time.
So in such kind of cases the advantage of our modification on stability does
not prevail.

Interestingly we do find our new modification has the advantage on accuracy
compared with the usual BSSN formulation.
The advantage is manifest on the gravitational radiation wave form calculation.
And this issue has been investigated in \cite{alic11,cao12}.
As we have mentioned, both the usual BSSN formulation and our new modification
are stable for simulations of QC1 and QC2.
For completeness, here we list the initial parameters for QC1 and QC2 here.
The initial positions of binary black hole for these two configurations are
$(0,\pm1.364,0)$ and $(0,\pm1.516,0)$.
The initial linear momentum are $(\mp0.286,0,0)$ and $(\mp0.258,0,0)$.
For both configurations the two black holes are spineless and identical with
puncture mass parameter $0.463$ and $0.47$ respectively.
We plot the resulted waveform in Fig.\ref{fig4}.
The waveform is measured by Newman-Penrose quantity $\Psi_4$.
Our calculation follows the description in \cite{czyy08}.
The figure shows that the two formulations give consistent waveform.
And the consistence convinces us the new modification does work well for
binary black hole simulations.
In particular, the formulation with our modification suppresses the numerical
error more effectively and produce a smoother waveform than the usual BSSN
formulation does during the ringdown phase.
We should emphasize that the boundary effects has no effect on the late-time
behavior of waveform. Since the outermost boundary is at $517.12$ and the
waveform is measured at $R=50$, therefore the noise in the waveform mainly
comes from the evolution itself.
\section{Summary}
We experiment with various modifications of the BSSN formulation and study
their effects on the stability of numerical evolution calculations of single
black hole.
Based on the modifications in \cite{yhbs02},
we enhance the unimodular determinant constraint with Eq.~(\ref{gzz}) and the
traceless extrinsic curvature constraint with Eq.~(\ref{Ayy}).
We further modify the evolution equation for the conformal connection functions
into Eq.~(\ref{dtGamma2}) by enhancing the linear term of ${\tilde\Gamma}^i$.
With an irreducible decomposition (\ref{decom}),
we replace the component ${\tilde\Gamma}^i_{\bf g}$ in the connection with
the ${\tilde\Gamma}^i$ and enhance the spatial secondary unimodular determinant
constraint ${\tilde\Gamma}^j{}_{ij}=0$ to form a new connection
$\boldsymbol{\mit\Gamma}{}^i{}_{ij}$ in Eq.~(\ref{newG}).
Meanwhile, the field equation of ${\tilde\Gamma}^i$ is adapted from
Eq.~(\ref{dtGamma2}) into Eq.~(\ref{dtGamma3}) for the new connection and to
ensure the suppression ability of numerical error in its linear term.
With the irreducible decomposition on ${\tilde\gamma}_{ij,k}$,
the field equation of the conformal metric is modified into Eq.~(\ref{dtg2}) by
adding the $\Gamma$-constraint.
The field equation of the conformal extrinsic curvature is modified into
Eq.~(\ref{modta2}) by combining the $\tilde A$-adjustment proposed in
\cite{ygsh02} and the irreducible decomposition on ${\tilde A}_{ij,k}$.

We found that these modifications on the BSSN formulation do show their
superiority on numerical error suppression compared with the earlier work when
applied to single Kerr-Schild black hole calculations and thus increase the
accuracy eventually.
When applied to the binary black hole calculations with the typical initial
data and without the Kreiss-Oliger dissipation, the modified BSSN formulation
gives a consistent and more accurate result than the conventional method.

Among these modifications, Eq.~(\ref{dtg2}) seems play a key role in stability
and thus the whole evolution is quite sensitive to the value of the parameter
$\sigma$.
We suspect that this modification, combined with the Ricci curvature in the
field equation of ${\tilde A}_{ij}$, could change the characteristic of
${\tilde\Gamma}^i$, and thus of the system.
However, a further analytic/numerical study is needed to have a better
understanding of the effect of this modification.

In this work, we only demonstrate the advantage of these modifications to
stability and accuracy of binary black hole simulations without any systematic
study on the optimal choice of the related parameters.
Thus we plan to  address this problem in more detail in a separate
upcoming work to increase the numerical accuracy of future binary black hole
simulations.
\section*{Acknowledgments}
This work was supported in part by the National Science Council
under the grants NSC98-2112-M-006-007-MY2 and NSC100-2112-M-006-005,
and by the National Center of Theoretical Sciences.
Z.~Cao was supported by the NSFC (No.~11005149). 
We are grateful to the National Center for High-performance Computing
for computer time and facilities, and
Academia Sinica Computing Center for providing computing resource. 
We also thank C. Soo for his helpful suggestions and comments.


\begin{thebibliography}{99}
\bibitem{pref05}
F.~Pretorius,
Phys.~Rev.~Lett.~\textbf{95}, 121101 (2005).

\bibitem{NR06}
M.~Campanelli, C.O.~Lousto, P.~Marronetti, and Y.~Zlochower,
Phys.~Rev.~Lett.~\textbf{96}, 111101 (2006);
J.G.~Baker, J.~Centrella, D.~Choi, M.~Koppitz, and J.~van Meter,
ibid., 111102 (2006).

\bibitem{Lehner_reviewNR01}
L.~Lehner,
Class.~Quant.~Grav.~{\bf 18}, R25 (2001).

\bibitem{reviewBBH}
J.~Centrella, J.G.~Baker, B.J.~Kelly, and J.R.~van~Meter,
Rev.~Mod.~Phys.~\textbf{82}, 3069 (2010);
I.~Hinder,
Class.~Quant.~Grav.~\textbf{27}, 114004 (2010);
S.T.~McWilliams,
ibid. \textbf{28}, 134001 (2011).

\bibitem{reviewBHNS}
T.W.~Baumgarte and S.L.~Shapiro, Phys.~Rept.~\textbf{376}, 41, (2003);
M.D.~Duez, Class.~Quant.~Grav.~\textbf{27}, 114002 (2010).

\bibitem{BHkicks}
M.~Campanelli, C.O.~Manuela, Y.~Zlochower, and D.~Merritt,
Phys.~Rev.~Lett.~\textbf{98}, 231102 (2007);
M.~Koppitz, D.~Pollney, C.~Reisswig, L.~Rezzolla, J.~Thornburg, P.~Diener, and E.~Schnetter, ibid. {\bf 99}, 041102 (2007);
C.O.~Lousto and Y.~Zlochower, Phys.~Rev.~D \textbf{77}, 044028 (2008);
C.O.~Lousto and Y.~Zlochower, ibid. \textbf{83}, 024003 (2011);
Y.~Zlochower, M.~Campanelli, and C.O.~Lousto, Class.~Quant.~Grav.~\textbf{28}, 114015 (2011);
C.O.~Lousto, Y.~Zlochower, M.~Dotti, and M.~Volonteri, arXiv:1201.1923.

\bibitem{JetBH}
J.C.~McKinney and R.D.~Blandford,
arXiv:0812.1060;
C.~Palenzuela, L.~Lehner, and S.L.~Liebling,
Science~\textbf{329}, 927 (2010);
P.~Moesta, D.~Alic, L.~Rezzolla, O.~Zanotti, and C.~Palenzuela,
Astrophys.~J.~\textbf{749}, L32 (2012).

\bibitem{JetNS}
L.~Rezzolla, B.~Giacomazzo, L.~Baiotti, J.~Granot, C.~Kouveliotou,
and M.A.~Aloy, Astrophys.~J.~\textbf{732}, L6 (2011).

\bibitem{lial06}
L.~Lindblom, M.A.~Scheel, L.E.~Kidder, R.~Owen, and O.~Rinne,
Class.~Quant.~Grav.~{\bf 23}, S447 (2006).

\bibitem{bral99}
O.~Brodbeck, S.~Frittelli, P.~Huebner, and O.A.~Reula,
J.~Math.~Phys.~{\bf 40}, 909 (1999);
C.~Gundlach, J.M.~Martin-Garcia, G.~Calabrese, and I.~Hinder,
Class.~Quant.~Grav.~{\bf 22}, 3767 (2005).

\bibitem{BSSN95_99}
M.~Shibata and T.~Nakamura,
Phys.~Rev.~D \textbf{52}, 5428 (1995);
T.W.~Baumgarte and S.L.~Shapiro,
ibid. {\bf 59}, 024007 (1999).

\bibitem{almb01}
M.~Alcubierre and B.~Br\"ugmann,
Phys.~Rev.~D \textbf{63}, 104006 (2001).

\bibitem{yhbs02}
H.J.~Yo, T.W.~Baumgarte, and S.L.~Shapiro,
Phys.~Rev.~D \textbf{66}, 084026 (2002).

\bibitem{ygsh02}
G.~Yoneda and H.~Shinkai,
Phys.~Rev.~D \textbf{66}, 124003 (2002).

\bibitem{kksh08}
K.~Kiuchi and H.~Shinkai,
Phys.~Rev.~D \textbf{77}, 044010 (2008).

\bibitem{ttys12}
T.~Tsuchiya, G.~Yoneda, and H.~Shinkai,
Phys.~Rev.~D \textbf{85}, 044018 (2012).

\bibitem{bjet12}
J.D.~Brown {\it et al.}, arXiv:1202.1038.

\bibitem{czyy08}
Z.~Cao, H.J.~Yo, and J.P.~Yu,
Phys.~Rev.~D \textbf{78}, 124011 (2008).

\bibitem{Z3Z4}
C.~Bona, T.~Ledvinka, and C.~Palenzuela, Phys.~Rev.~D \textbf{66}, 084013(2002);
C.~Bona, T.~Ledvinka, C.~Palenzuela, and M.~Zacek, ibid. \textbf{67}, 104005 (2003).

\bibitem{LaPS02}
P.~Laguna and D.~Shoemaker,
Class.~Quant.~Grav.~\textbf{19}, 3679 (2002).

\bibitem{OVEH97}
Y.N.~Obukhov, E.J.~Vlachynsky, W.~Esser, and F.W.~Hehl,
arXiv:gr-qc/9705039.

\bibitem{SpeU07}
U.~Sperhake, Phys.~Rev.~D \textbf{76}, 104015 (2007).

\bibitem{mrhs99}
R.A.~Matzner, M.F.~Huq, and D.~Shoemaker,
Phys.~Rev.~D \textbf{59}, 024015 (1999).

\bibitem{chas92}
S.~Chandrasekhar, {\it The Mathematical Theory of Black Holes}
(Oxford University Press, New York, 1992).

\bibitem{edda58}
A.E.~Eddington, Nature {\bf 113}, 192 (1924);
D.~Finkelstein, Phys.~Rev. \textbf{110}, 965 (1958).

\bibitem{tichy04}
W.~Tichy and B.~Br\"ugmann,
Phys.~Rev.~D \textbf{69}, 024006 (2004).

\bibitem{lorene}
E.~Gourgoulhon, P.~Grandcl\'ement, K.~Taniguchi, J.~Marck, and S.~Bonazzola,
Phys.~Rev.~D \textbf{63}, 064029 (2001);
K.~Taniguchi and E.~Gourgoulhon,
ibid. \textbf{66}, 104019 (2002); ibid. \textbf{68}, 124025 (2003).

\bibitem{loreneweb}
http://www.lorene.obspm.fr/

\bibitem{IDlecture}
G.B.~Cook,
Living~Rev.~Relativity \textbf{3}, 5 (2000);
E.~Gourgoulhon,
J.~Phys.:~Conf.~Ser.~\textbf{91}, 012001 (2007).

\bibitem{CTT1}
A.P.~Lichnerowicz,
J.~Math.~Pures~Appl.~\textbf{23}, 37 (1944)

\bibitem{CTT2}
J.W.~York,
Phys.~Rev.~Lett.~\textbf{26}, 1656 (1971);
J.W.~York,
J.~Math.~Phys.~\textbf{14}, 456 (1973);
N.~\'O~Murchadha and J.W.~York,
Phys.~Rev.~D \textbf{10}, 428 (1974).

\bibitem{BJYJ80}
J.M.~Bowen and J.W.~York,
Phys.~Rev.~D \textbf{21}, 2047 (1980).

\bibitem{BDLR63}
D.R.~Brill and R.W.~Lindquist,
Phys.~Rev.~\textbf{131}, 471 (1963).

\bibitem{brandt97}
S.~Brandt and B.~Br\"ugmann,
Phys.~Rev.~Lett.~\textbf{78}, 3606 (1997).

\bibitem{GEGB02}
E.~Gourgoulhon, P.~Grandcl\'ement, and S.~Bonazzola,
Phys.~Rev.~D \textbf{65}, 044020 (2002);
P.~Grandcl\'ement, E.~Gourgoulhon, and S.~Bonazzola,
ibid., 044021 (2002).

\bibitem{two_puncture}
M.~Ansorg, B.~Bru\"gmann, and W.~Tichy,
Phys.~Rev.~D \textbf{70}, 064011 (2004).

\bibitem{3BH}
Z.~Cao, J.P.~Yu, C.Y.~Lin, S.~Bai, and H.J.~Yo,
arXiv:1203.6185

\bibitem{bmss95}
C.~Bona, J.~Masso, E.~Seidel, and J.~Stela,
Phys.~Rev.~Lett.~\textbf{75}, 600 (1995).

\bibitem{aaea99}
A.~Arbona {\it et al.},
Phys.~Rev.~D \textbf{60}, 104014 (1999).

\bibitem{amea00}
M.~Alcubierre {\it et al.},
Phys.~Rev.~D \textbf{62}, 044034 (2000).

\bibitem{meter06}
J.~Meter, J.~Baker, M.~Koppitz and, D.~Choi,
Phys.~Rev.~D \textbf{73}, 124011 (2006).

\bibitem{balakrishna96}
J.~Balakrishna, G.~Daues, E.~Seidel, W.~Suen, M.~Tobias, and E.~Wang,
Class.~Quant.~Grav.~\textbf{13}, L135 (1996).

\bibitem{cao10}
P.~Galaviz, B.~Br\"ugmann, and Z.~Cao,
Phys.~Rev.~D \textbf{82}, 024005 (2010).

\bibitem{cao12}
Z.~Cao and D.~Hilditch, arXiv:1111.2177.

\bibitem{baker02}
J.~Baker, M.~Campanelli, C.~Lousto, and R.~Takahashi,
Phys.~Rev.~D \textbf{65}, 124012 (2002).

\bibitem{alic11}
D.~Alic, C.~Bona-Casas, C.~Bona, L.~Rezzolla, and C.~Palenzuela,
arXiv:1106.2254.

\end{thebibliography}
\end{document}